\documentclass[aps,prb,10pt,twocolumn,amsmath,amssymb,showpacs, superscriptaddress]{revtex4-1}

\usepackage{graphicx}% Include figure files
\usepackage{amsmath}
\usepackage{mathbbol}
\usepackage{subfigure}

\def\beq{\begin{equation}}
\def\eeq{\end{equation}}
\def\bea{\begin{eqnarray}}
\def\eea{\end{eqnarray}}

\def\mos{{MoS$_2$}}
\def\k{{\mathbf{k}}}
\def\K{{\mathbf{K}}}
\def\ef{\varepsilon_F}
\def\de{\delta\varepsilon}

\newcommand{\eq}[1]{Eq.~(\ref{eq:#1})}
\newcommand{\eqs}[2]{Eqs.~(\ref{eq:#1}) and~(\ref{eq:#2})}
\newcommand{\equ}[1]{Equation~(\ref{eq:#1})}

\begin{document}

\title{Valley Hall effect in disordered monolayer 
	     MoS$_2$ from first principles}

\author{Thomas Olsen} \email{tolsen@fysik.dtu.dk} \affiliation{Centro
  de F{\'i}sica de Materiales, Universidad del Pa{\'i}s Vasco, 20018
  San Sebast{\'i}an, Spain} \affiliation{Center for Atomic-Scale
  Materials Design, Department of Physics, Technical University of
  Denmark}

\author{Ivo Souza} \email{ivo_souza@ehu.es} \affiliation{Centro de
  F{\'i}sica de Materiales, Universidad del Pa{\'i}s Vasco, 20018 San
  Sebast{\'i}an, Spain} \affiliation{Ikerbasque Foundation, 48013
  Bilbao, Spain}

\date{\today}

\begin{abstract}
  Electrons in certain two-dimensional crystals possess a pseudospin
  degree of freedom associated with the existence of two inequivalent
  valleys in the Brillouin zone. If, as in monolayer MoS$_2$, inversion
  symmetry is broken and time-reversal symmetry is present, equal and
  opposite amounts of $k$-space Berry curvature accumulate in each of
  the two valleys. This is conveniently quantified by the integral of
  the Berry curvature over a single valley - the valley Hall
  conductivity.  We generalize this definition to include
  contributions from disorder described with the supercell approach,
  by mapping ("unfolding") the Berry curvature from the folded
  Brillouin zone of the disordered supercell onto the normal Brillouin
  zone of the pristine crystal, and then averaging over several
  realizations of disorder. We use this scheme to study from
  first-principles the effect of sulfur vacancies on the valley Hall
  conductivity of monolayer MoS$_2$. In dirty samples the intrinsic
  valley Hall conductivity receives gating-dependent corrections that
  are only weakly dependent on the impurity concentration, consistent
  with side-jump scattering and the unfolded Berry curvature can be interpreted
  as a $k$-space resolved side-jump. At low impurity concentrations
  skew scattering dominates, leading to a divergent valley Hall
  conductivity in the clean limit. The implications
  for the recently-observed photoinduced anomalous Hall effect are
  discussed.
\end{abstract}
\maketitle

\section{Introduction}
\begin{figure}[tb]
    \includegraphics[width=4.0 cm]{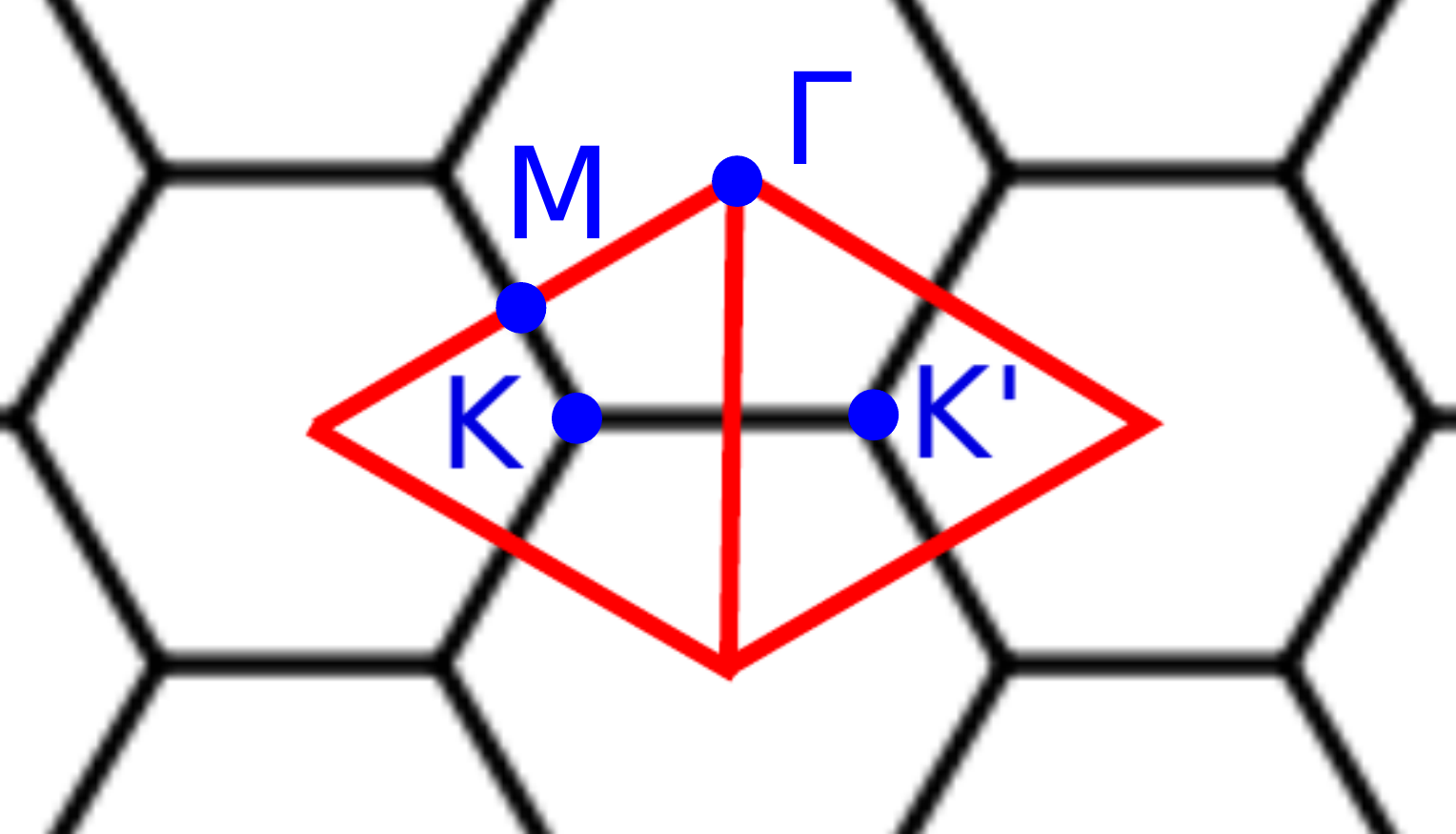}
    \caption{(Color online). Brillouin zone of monolayer MoS$_2$,
      partitioned into two triangular valleys delimited by lines of
      vanishing Berry curvature, drawn in red.  The two valleys are
      centered at the high-symmetry points $K$ and $K'$.}
\label{fig:valley}
\end{figure}

Monolayers of MoS$_2$ and related transition-metal dichalcogenides
(TMDs) have recently become the subject of intense investigation, due
in part to the possibility of manipulating the so-called ``valley''
degree of freedom.\cite{xu-natphys14} These materials have the
symmetry of a honeycomb structure with a staggered sublattice, thus
lacking an inversion center. The bandstructure exhibits a direct gap
at the two inequivalent valleys centered at the high-symmetry
points~$K$ and $K'=-K$ in the Brillouin zone (see
Fig. \ref{fig:valley}), where the topmost valence bands are are
primarily composed of transition-metal $d$~states.\cite{mak_photo}
Time-reversal symmetry, which takes $\mathbf{k}$ into $-\mathbf{k}$ and therefore maps
one valley onto the other, dictates that states in a given band at $K$
and $K'$ carry antiparallel angular momenta. This inspired Xiao {\it
  et al.} to propose using circularly-polarized light as a means of
selectively exciting carriers from a particular
valley.\cite{xiao_graph, xiao_valley} The effect was rapidly confirmed
experimentally, by demonstrating that excitation with
circularly-polarized light results in polarized
fluorescence.\cite{zeng, mak_optical_valley, cao}

The broken inversion symmetry in monolayer MoS$_2$ induces a nonzero
Berry curvature on the Bloch bands (in contrast, the Berry curvature
vanishes identically for bilayer and bulk \mos, both of which are
centrosymmetric). The Berry curvature is defined in terms of the
cell-periodic Bloch states as
\beq
\label{eq:curv-n}
\Omega_{n,xy}(\k)=-2\text{Im}\sum_n\,
\langle\partial_{k_x}u_{n\k}|\partial_{k_y}u_{n\k}\rangle\,,
\eeq
and it modifies the current response to an applied electric field by
adding an ``anomalous velocity'' term to the semiclassical equations
of motion.\cite{xiao-rmp10} A well-known consequence is the anomalous
Hall effect (AHE) in magnetic materials, where the Berry curvature is
induced by broken time-reversal symmetry. The intrinsic, 
%
%\ism{I have introduced this notation to resolve a notational
%  inconsistency that existed in the previous version of the manuscript
%  between the left-hand-sides of Eqs.~\eqref{eq:vhc-dirac}
%  and~\eqref{intrinsic}.}
%
part of the anomalous
Hall conductivity (AHC) is given by the Brillouin zone (BZ) integral
of the Berry curvature summed over the occupied
states,\cite{xiao-rmp10,nagaosa_review}
\bea
\label{eq:curv-k}
\Omega_{xy}({\bf k})&=&\sum_n\,f_{n{\bf k}} \Omega_{n,xy}({\bf k})\\
\label{eq:ahc}
\sigma_{xy}^{0}&=&-\frac{e^2}{h}\int_{\rm BZ}\,\frac{d^d
  k}{(2\pi)^{d-1}}\Omega_{xy}({\bf k})\,,
\eea
where $f_{n{\bf k}}$ is the occupation factor and $d$ is the
dimensionality. Here, the superscript 0 denoted that it is the 
intrinsic part of the AHC. For $d=2$ the AHC has units of conductance~(S), and
for $d=3$ it has units of conductivity~(S/cm).
The broken inversion symmetry in monolayer MoS$_2$ induces a nonzero
Berry curvature on the Bloch bands (in contrast, the Berry curvature
vanishes identically for bilayer and bulk \mos, both of which are
centrosymmetric). The Berry curvature is defined in terms of the
cell-periodic Bloch states as
Monolayer \mos\, is nonmagnetic, and the presence of time-reversal
symmetry implies the relation\cite{xiao-rmp10}
\beq
\label{eq:curv-tr}
\Omega_{xy}(-\mathbf{k})=-\Omega_{xy}(\mathbf{k})\,.
\eeq
Thus, equal and opposite amounts of Berry curvature accumulate in the
two valleys, resulting in a cancellation of the valley Hall currents
and a vanishing AHC. Time-reversal symmetry can however be broken by
illuminating the sample with circularly-polarized light, leading to a
{\it photoinduced} AHE. The valley-selective photoexcitation creates a
carrier imbalance which in turn removes the exact cancellation between
the Hall currents in the two valleys. This so-called {\it valley Hall
  effect} was first discussed for graphene systems with broken
inversion symmetry,\cite{xiao_graph} and later for monolayer
\mos.\cite{xiao_valley} The effect was subsequently measured by Mak
{\it at al.} in transistors of MoS$_2$ monolayers.\cite{mak_valley}

Compared to the conventional AHE in ferromagnetic metals, the
theoretical modeling of the photoinduced AHE in TMDs poses the
additional challenge that the AHC should in principle be calculated
for a \textit{nonequilibrium} photoexcited state, but to our
knowledge, such a calculation has not yet been attempted.  Instead, an
approximate but more tractable approach is often
used.\cite{xiao_graph,xiao_valley,mak_valley} The idea is to introduce
an auxiliary quantity $\sigma_{xy}^V$, which we will call the {\it
  valley Hall conductivity} (VHC), defined as the integral of the
Berry curvature over a single valley domain in the BZ. For example,
the intrinsic VHC of the valley centered at $K$ in Fig.~\ref{fig:valley} is
\beq
\label{eq:vhc-KL}
\sigma_{xy}^{0,K}(\ef)=
-\frac{e^2}{h}\int_K\frac{d^2k}{2\pi}\Omega_{xy}(\mathbf{k})\,.  
\eeq
and similarly for the valley centered at $K'$. (The demarcation
of the two valley domains will be discussed further in
Sec.~\ref{sec:elec-struct}. \equ{vhc-KL} depends on the Fermi
level $\ef$ through the occupation factors in \eq{curv-k}.)  The
photoinduced AHC $\sigma_{xy}$
%
%\ism{TO DO: make sure that $\Delta\sigma_{\rm H}$ is replaced with
%  $\sigma_{xy}$ everywhere in the manuscript (including figures).}
%
is then approximated by the sum of the VHCs in the two valleys,
positing a Fermi-level shift $\de$ between them to mimic the effect of
the valley-selective photoexcitation,
\beq
\label{eq:photo-ahc}
% \Delta\sigma_{\rm H}(\ef,\de)=
% \sigma_{\rm VH}^{K}(\ef+\de)+\sigma_{\rm VH}^{K'}(\ef)\,.  
\delta\sigma_{xy}(\ef,\de)=\sigma_{xy}^K(\ef+\de)+\sigma_{xy}^{K'}(\ef)\,.
\eeq
When $\de=0$ the AHC vanishes, and a nonzero
$\delta\sigma_{xy}$ appears when $\de\not= 0$. This approach,
also allows for a direct comparison with model calculations
without considering the details of how the carrier imbalance between
valleys is generated. 
%

%\ism{Added, to drive home the point you mentioned in your reply to
%  what was then marginpar~[9].}
%
This will be the basic approach taken in the present work. We
  have suppressed the superscript 0 from this last equation to
  emphasize that it remains valid when the non-intrinsic contributions which
  we will now discuss are taken into account.

Impurities are always present in real samples, and their {\it
  extrinsic} contributions to the photoinduced AHE in TMDs should be
taken into account alongside the intrinsic response described by
\eqs{vhc-KL}{photo-ahc}. This is well known in the
context of the AHE in ferromagnetic metals, where historically two
types of extrinsic contributions have been considered - \textit{side
  jump} and \textit{skew scattering}.\cite{nagaosa_review} In a
simplified picture, the side-jump effect originates in the anomalous
velocity that a wave packet may acquire as it moves through an impurity
potential,
%
%\ism{Do you want to remove this? We should probably clarify, either
%  here or later, that while SOC, either in the lattice potential
%  or/and in the impurity potential, is needed for skew-scattering and
% side jump in ferromagnetic metals, it is not strictly needed in
%  TMDs.  A good place to mention it might be in the same paragraph
%  where we comment that contrary to ferromagnetic metals, the
%  intrinsic Berry curvature in TMDs does not require SOC. We could
%  then add that likewise, the impurity contributions (skew-scattering
%  and side-jump) do not require SOC.}
%
 while skew scattering arises from
the chiral part of a standard transition-rate expression.  With some
effort, both contributions can be incorporated into the semiclassical
Boltzmann-transport framework.\cite{sinitsyn_coordinate}

The correspondence between the semiclassical treatment of the AHC and
a fully quantum-mechanical (Kubo-Streda) calculation based on a
perturbative expansion in powers of the disorder strength was
carefully worked out in Ref.~\onlinecite{sinitsyn_link}. It became
clear from that analysis that not all terms fall distinctly into
either of the above physical interpretations of extrinsic
contributions to the AHC, and for many purposes it is more practical
to base the distinction on the scaling with impurity
concentration.\cite{nagaosa_review} According to this viewpoint
skew-scattering is defined as the part of the AHC which scales
inversely with the impurity concentration, while the part which is
independent of the impurity concentration has both intrinsic and
side-jump components. Although the intrinsic contribution is sharply
defined theoretically in terms of the electronic structure of the
pristine crystal by \eq{ahc}, experimentally it is not known how to
separate it from the side-jump part. Note that the anomalous Hall
response of pristine samples at low temperatures is dominated by
skew-scattering, with the intrinsic contribution only becoming
significant in moderately resistive samples (where it competes with
side-jump scattering). This analysis, originally developed for the AHC
in ferromagnetic metals, carries over to the VHC and photoinduced AHC
in TMDs.

It is well established that sulfur vacancies constitute the main
source of disorder in MoS$_2$.\cite{zhou_defects, liu_defects,
  ma_defects, zhou_defects1, wei_defects, asl_defects} The formation
energies and thermodynamics of these defects have been thoroughly
studied,\cite{komsa} but their influence on transport and optical
properties remains largely unexplored. Modeling the effects of
disorder from first-principles is a challenging task in general, but
there are noteworthy examples where the AHC in ferromagnetic materials
has been calculated using the coherent potential
approximation;\cite{velicky, faulkner, butler, ebert, blugel_skew,
  kudr} also, an \textit{ab initio} implementation of the side-jump
contribution to the AHC has been carried out assuming scattering
centers with delta-function potentials.\cite{jurgen}

In this work, we develop a computational scheme that allows us to
include in a realistic manner the effect of impurities in the
calculation of the VHC and of the photoinduced AHE in TMDs. In a first
step, we perform several supercell calculations at the desired
impurity concentration, corresponding to different realizations of
disorder. In order to carry out the calculations efficiently while
maintaining first-principles-like accuracy, we construct effective
Hamiltonians in a Wannier-function basis, starting from density
functional theory calculations on smaller
cells.\cite{berlijn_effective} Recall that the definition of the VHC
in \eq{vhc-KL} requires identifying the individual valley
domains in the BZ where the Berry curvature is to be integrated. It is
not clear \textit{a priori} how to do so in the context of a supercell
calculation, since the electronic states cannot be labeled by
wavevectors in the normal BZ of Fig.~\ref{fig:valley}.  In order to
overcome this difficulty, in a second step we use a ``BZ unfolding''
technique\cite{ku_unfolding} to map the results of the supercell
calculation onto the normal BZ of the pristine crystal. More
precisely, we express the AHC of each disordered supercell
configuration as an integral of the {\it supercell} Berry curvature
over the folded BZ, and then unfold the Berry curvature onto the
normal BZ according to the prescription of
Ref.~\onlinecite{bianco_unfolding}. Having done that, the VHC
(including the contributions from disorder) can then be obtained by
integrating the {\it unfolded} Berry curvature over a single valley
domain in Fig.~\ref{fig:valley}, and averaging the result over several
realizations of disorder.

We have used the above first-principles-based methodology to study the
influence of sulfur vacancies on the VHC of \mos as well as the 
photoinduced AHC, which is the quantity measured in experiments. The calculated VHC
as a function of defect concentration was compared with model
calculations where the valence and conduction-band edges in each
valley are described by a massive Dirac Hamiltonian with a random
distribution of delta-function scatterers.\cite{sinitsyn_link}

The paper  is organized  as follows. We  start Sec.  \ref{pristine} by
reviewing  some   basic  features  of  the   electronic  structure  of
MoS$_2$. We  then evaluate  the intrinsic VHC  [\eq{vhc-KL}] and
photoinduced  AHC  [\eq{photo-ahc}]  for  the  massive  Dirac
Hamiltonian  without   disorder,  and  carry  out   the  corresponding
\textit{ab initio} calculations for pristine MoS$_2$. Our main results
are  presented in  Sec. \ref{disordered},  where disorder  effects are
included in  the calculation of  the VHC,  both for the  massive Dirac
Hamiltonian and  for \mos\, with  sulfur vacancies.  The two  types of
calculations  are found  to be  in reasonable  agreement, and  we then
proceed to calculate  the photoinduced AHC for  the disordered massive
Dirac model  as a  function of gating  voltage and  Fermi-level shift,
finding good agreement with  the experimental measurements. Our
conclusions are  summarized in Sec.~\ref{sec:conclusions},  and the
appendices present the details of the \textit{ab initio} calculations,
the BZ unfolding method, and the effective-Hamiltonian methodology.

\section{Pristine MoS$_2$}\label{pristine}
\subsection{Energy bands and Berry curvature
% Electronic structure
}
\label{sec:elec-struct}
\begin{figure}[tb]	
    \includegraphics[width=8.0 cm]{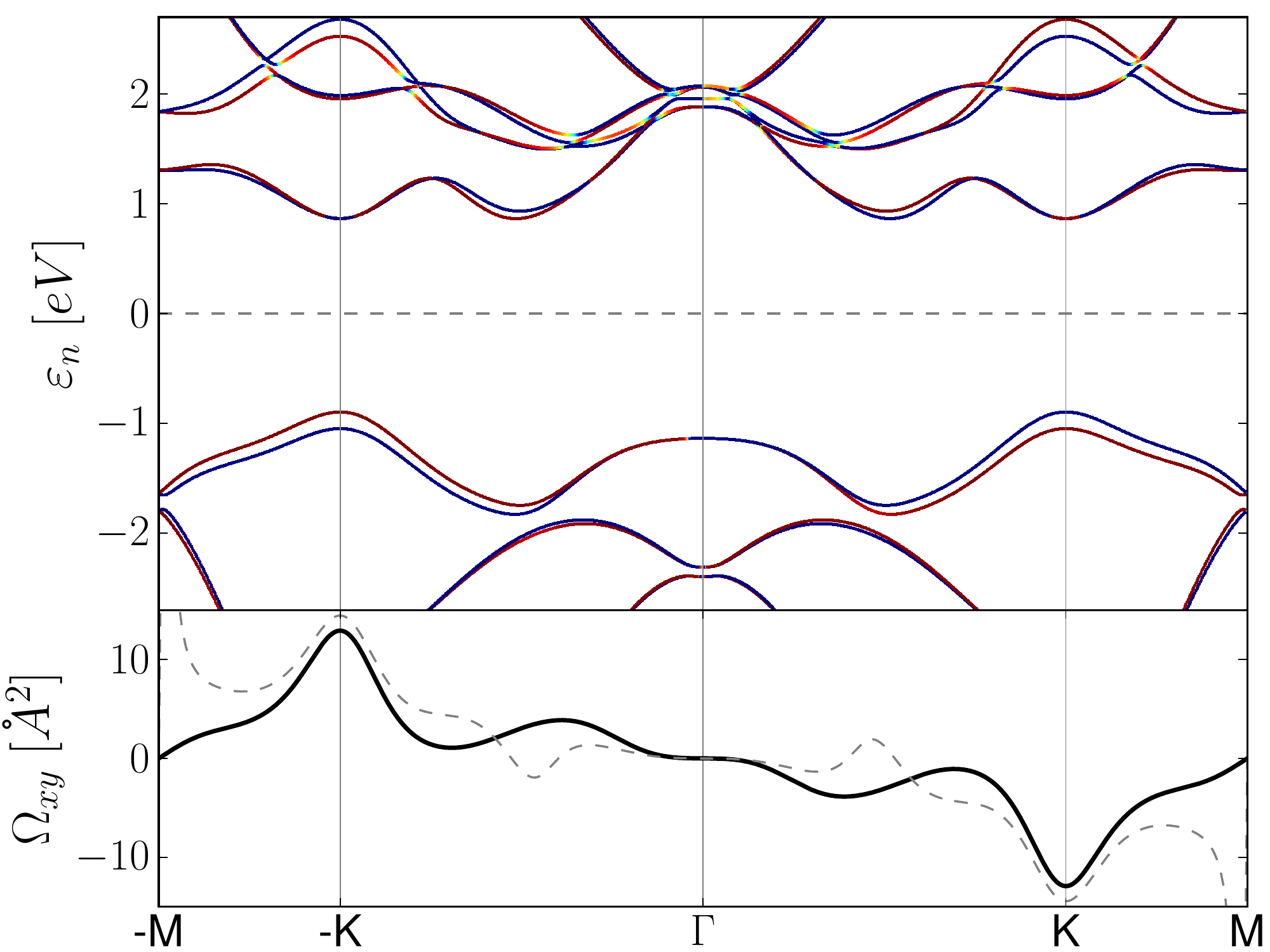}
    \caption{(Color online). Top: Calculated bandstructure of
      monolayer MoS$_2$. Energies are measured from the Fermi
        level, and the bands are color-coded by the spin expectation
      value $\langle S_z \rangle$, with red corresponding to spin up
      and blue to spin down. Bottom: Berry curvature summed
      over the valence bands [\eq{curv-k}]. The dashed line is the 
      Berry curvature evaluated from the two topmost valence bands.}
\label{fig:pristine_band_curv}
\end{figure}

{\it Ab initio} density-functional theory calculations were
  carried out for monolayer \mos\, as described in
  Appendix~\ref{sec:abinitio}.  The calculated Kohn-Sham
%
%\ism{Expanded a bit opening sentence, in view of email comments. OK?}
%
energy bands are shown in the upper panel of
Fig.~\ref{fig:pristine_band_curv}, color-coded by the spin component
$\langle S_z\rangle$ orthogonal to the layer.  The minimum direct gap
is situated at $K$ and $K'=-K$, with a value of $\sim1.7$~eV.
Away from the time-reversal-invariant points~$\Gamma$ and~${\rm
  M}$ the spin degeneracy is split by the combination of broken
inversion symmetry and spin-orbit coupling (the degeneracy is actually
protected along the entire $\Gamma$-${\rm M}$ line by mirror
symmetry).  The two topmost valence bands exhibit a maximum spin-orbit
splitting of $\sim0.15$~eV at $K$ and $K'$, where $S_z$ is a good
quantum number and Kramers-degenerate partners have opposite spin
character: $\varepsilon_{K\uparrow}=\varepsilon_{K'\downarrow}$.

The lower panel of Fig. \ref{fig:pristine_band_curv} shows the Berry
curvature summed over the valence bands, \eq{curv-k}. In agreement
with \eq{curv-tr}, $\Omega_{xy}$ is an odd function of $\k$.  Its
magnitude peaks at the minimum-gap points~$K$ and~$K'$, where the sign
is dictated by the dominant contributions coming from the topmost
valence bands. At~$K$ those bands are mainly composed of molybdenum
$d$-states with $m_l=2$; according to the optical selection
rules\cite{cao} those states can be excited with left-handed polarized
light.
%
%\ism{I will let you decide what to do here.}
%
The Berry curvature has a secondary peak between~$\Gamma$ and~$K$;
there, a pair of lower-lying valence bands approaches the topmost
ones, and also contributes significantly to the Berry curvature.

It should be noted that because the Berry curvature is induced by the
broken spatial inversion, it is not directly related to the spin-orbit
splitting evident in Fig.~\ref{fig:pristine_band_curv}; in fact,
$\Omega_{xy}$ is practically unaltered if the spin-orbit interaction
is switched off. (This is in sharp contrast to the Berry curvature
induced by broken time-reversal symmetry in ferromagnetic metals,
which vanishes in the absence of
spin-orbit coupling.\cite{xiao-rmp10,nagaosa_review}) Likewise, the extrinsic 
scattering contributions do not rely on spin-orbit coupling. Thus one can use a
spinless model such as the massive Dirac Hamiltonian of
Sec.~\ref{sec:dirac-model} to describe the band edges and valley Berry
curvature in \mos; spin is then accounted for by inserting a factor of
two in the calculated $\Omega_{xy}$.

%\ism{Worked a bit more on this paragraph.}
%
The Hamiltonian of monolayer MoS$_2$ is invariant under
reflection across the vertical planes containing the lines that
connect a Mo~atom to the neighboring S~atoms. The corresponding
symmetry elements in reciprocal space are the $\Gamma$-${\rm M}$
mirror lines. The Berry curvature transforms like a magnetic field
in reciprocal space.\cite{xiao-rmp10} In particular, the component
$\Omega_{xy}=\Omega_z$ is odd under reflection across the
$\Gamma$-${\rm M}$ lines, and hence it vanishes along those lines,
which form the boundaries between the two valleys in
Fig.~\ref{fig:valley}. This allows us to uniquely define the
intrinsic VHC according to \eq{vhc-KL}.

\subsection{Intrinsic Valley Hall conductivity}

%\ism{Tentatively commented out the subsubsection that was here, since
%  the material is now covered in the Introduction. Check if there is
%  any text that you want to salvage...}

\subsubsection{Massive Dirac model}\label{sec:dirac-model}

The valence and conduction-band edges of a single valley of
MoS$_2$ and related materials are often modeled by the massive Dirac
Hamiltonian.\cite{xu-natphys14,xiao_graph, xiao_valley} The Hamiltonian 
for the $K$~valley in Fig. \ref{fig:pristine_band_curv} reads

\begin{align}\label{H_dirac}
 H(\k)=\hbar v(-k_x\sigma_x+k_y\sigma_y)+\Delta\sigma_z\,,
\end{align}
where $\k=(k_x,k_y)$ is measured from the valley center $K$,
  $\sigma_i$ are the Pauli matrices, and $\Delta$ is the mass
  parameter. The energy eigenvalues and Berry curvature are
\bea
\label{eq:eig-dirac}
 \varepsilon_{\pm}(\k)&=&\pm\sqrt{\Delta^2+v^2\hbar^2k^2}\\
\label{eq:curv-dirac}
 \Omega_{\pm,xy}(\k)&=&
\pm\frac{\Delta v^2\hbar^2}{2(\Delta^2+v^2\hbar^2k^2)^{3/2}}\,,
\eea
where  $k^2=k_x^2+k_y^2$.  In  the   case  of  the  $K'$  valley
\eq{eig-dirac} remains unchanged, while \eq{curv-dirac} flips sign.

If the Fermi level lies in the conduction band ($\ef>\Delta$),
the VHC becomes
%
%\ism{This is an equation for the physical VHC, so a factor of two
%  should be inserted explicitly to account for spin.}
%
\begin{align}\label{eq:vhc-dirac}
\sigma_{xy}^{0,K}&=
-\frac{2e^2}{h}\bigg[\int\frac{d^2k}{2\pi}\Omega_-(k)
+\int_{|k|<k_F}\frac{d^2k}{2\pi}\Omega_+(k)\bigg]\notag\\
&=\frac{e^2}{2h}-\frac{e^2\Delta}{2h}\Big(\frac{1}{\Delta}-
\frac{1}{\varepsilon_F}\Big).
\end{align}
The valley and spin degrees of freedom can be included by coupling
four copies of this model corresponding to spin and valley degrees of
freedom.
%
%\ism{I would remove this sentence and replace it with a sentence like,
%  {\it ``where an overall factor of two was inserted to account for
%    spin.''}}
%
 Referring to the band structure of MoS$_2$ we find that in
the vicinity of the valleys the massive Dirac model provides a good
fit if we take $\Delta=0.86$ eV. The velocity $v$ is calculated as
$v^2=\Delta/m^*_e$, where $m^*_e=0.4m_e$ is the effective mass of the
conduction band valley. 

For the ungated case where $\varepsilon_F$ is in the gap, the model gives an
intrinsic VHC
%
%\ism{Should it be VHC here?}
%
of $e^2/2h$. The deviation from that result measures the contribution from lower lying valence bands and 
the crystal potential, which gives rise to non-hyperbolic bands away from
the top of the valleys. It should be noted though, that a Chern
insulating system with an inversion center will retain the exact
value $\sigma_{xy}^{0,K}=e^2/2h$ when the crystal
potential is included, because symmetry implies
$\Omega(\k)=\Omega(-\k)$ and the topology implies
$\sigma_{xy}^{0}=2\sigma_{xy}^{0,K}=e^2/h$.\cite{haldane, kane_review}
%
%\ism{The previous sentence is a digression, and it interrupts the flow
%  of the text. Maybe it belongs in an explanatory footnote?}
%
For MoS$_2$ we find $\sigma_{xy}^{0,K}=0.71e^2/h$ and since each
spin channel contributes an equal amount of curvature,
%
%\ism{Again, in my opinion this factor of two due to spin should
%  already be included in the definition of the VHC.}
%
this corresponds to 71\% of the result for the massive Dirac
Hamiltonian.

As discussed in the introduction, measuring the valley Hall
effect requires generating a carrier imbalance between the two
valleys. Since experimental realizations usually involves a gate, we
include an overall Fermi level $\varepsilon_F$
%
%\ism{I feel that this discussion could be written more clearly.
%  Energies are measured relative to the equilibrium Fermi level
%  without gating, as in Fig.~\ref{fig:pristine_band_curv}, and
%  $\ef\not=0$ indicates an {\it overall} (same for both valleys)
%  Fermi-level offset caused by gating, whereas $\de$ is an offset of
%  the Fermi level at $K$ {\it relative} to the Fermi level at $K'$,
%  due to the presence of circularly-polarized light. I feel that the
%  best place to introduce these concepts would be around
%  \eqs{vhc-KL}{photo-ahc}. In fact, it might be
%  sufficient to expand the parenthesis below \eq{vhc-KL} to
%  indicate that we have indicated an explicit dependence on $\ef$
%  because experimentally $\ef$ is varied by gating.}
%
such that the Fermi
level in one valley is $\varepsilon_F$ and in the other valley it is
$\varepsilon_F+\delta\varepsilon$ due to the presence of polarized
light.  If we assume left-handed polarized light, we can then
calculate the photoinduced AHC as a function of $\delta\varepsilon$ in
the $K$~valley, which will depend on the strength of the optical
perturbation in an experimental setup. Combining 
\eqs{photo-ahc}{vhc-dirac} we obtain for $\ef>\Delta$,
%
%\ism{Commented out the middle equality, which corresponds to
%  \eq{photo-ahc}. Insert a factor of two for spin? What about
%  in the subsequent equations in this section?}
%
\begin{align}
\delta\sigma_{xy}^{0}(\ef,\de)
=-\frac{e^2\Delta}{2h}\Big(\frac{1}{\ef}-\frac{1}{\ef+\de}\Big)\,.
\end{align}
We can also express this quantity in terms
of carrier imbalance $\delta n_c$ between the conduction-band 
edges in the two valleys. For a single valley
\begin{align}
  n_c(\ef)=\frac{\pi
    k_F^2}{(2\pi)^2}=\frac{\varepsilon_F^2-\Delta^2}{4\pi v^2\hbar^2},
\end{align}
so that
%
%\ism{Is the middle expression in \eq{dnc} correct? It seems the same
%  as in the previous equation! Shouldn't the first equality in
%  \eq{dnc} be $\delta n_c\equiv n_c(\ef+\de)-n_c(\ef)$?}
%
\begin{align}
\label{eq:dnc}
\delta n_c=\frac{2\varepsilon_F\delta\varepsilon+\delta\varepsilon^2}{4\pi v^2\hbar^2}.
\end{align}
For $\delta\varepsilon\ll\varepsilon_F$ we then get
\begin{align}\label{ahc_linear}
\delta\sigma_{xy}^{0}\approx
-\frac{e^2}{h}\frac{\Delta}{\varepsilon_F^2}\delta\varepsilon\approx-\frac{2\pi e^2v^2\hbar\Delta}{\varepsilon_F^3}\delta n_c.
\end{align}
Thus, in this limit the
photoinduced AHC becomes linear in both the small energy shift
$\de$ and in the small carrier imbalance $\delta n_c$. An expression
similar to this one was derived in Ref. \onlinecite{mak_valley}, but with 
the Fermi level in the gap. That expression can be obtained by setting 
$\varepsilon_F=\Delta$ in Eq. \eqref{ahc_linear}. As will be shown in 
Sec. \ref{sec:photo_ind}, the position of the Fermi level is minuscule
for the intrinsic as well as the sidejump contributions, but can have a large effect
on the skew scattering contribution, which becomes dominant for
clean samples.

%
%\ism{Somewhere in this section (probably around here at the end) we
%  should make contact with the literature. After all this is a review,
%  or re-derivation, of know results, correct? If so we should conclude
%  with a sentence where we say that the expression we obtained agrees
%  with the one given in this or that paper (I guess
%  Ref.~\onlinecite{mak_photo}, but maybe others?) We may also want to
%  mention that the limit where Eq.~\eqref{ahc_linear} is valid
%  corresponds to the experimentally relavant regime.}

%\tnewpage

\subsubsection{First-principles calculations}

\begin{figure}[tb]
    \includegraphics[width=8.0 cm]{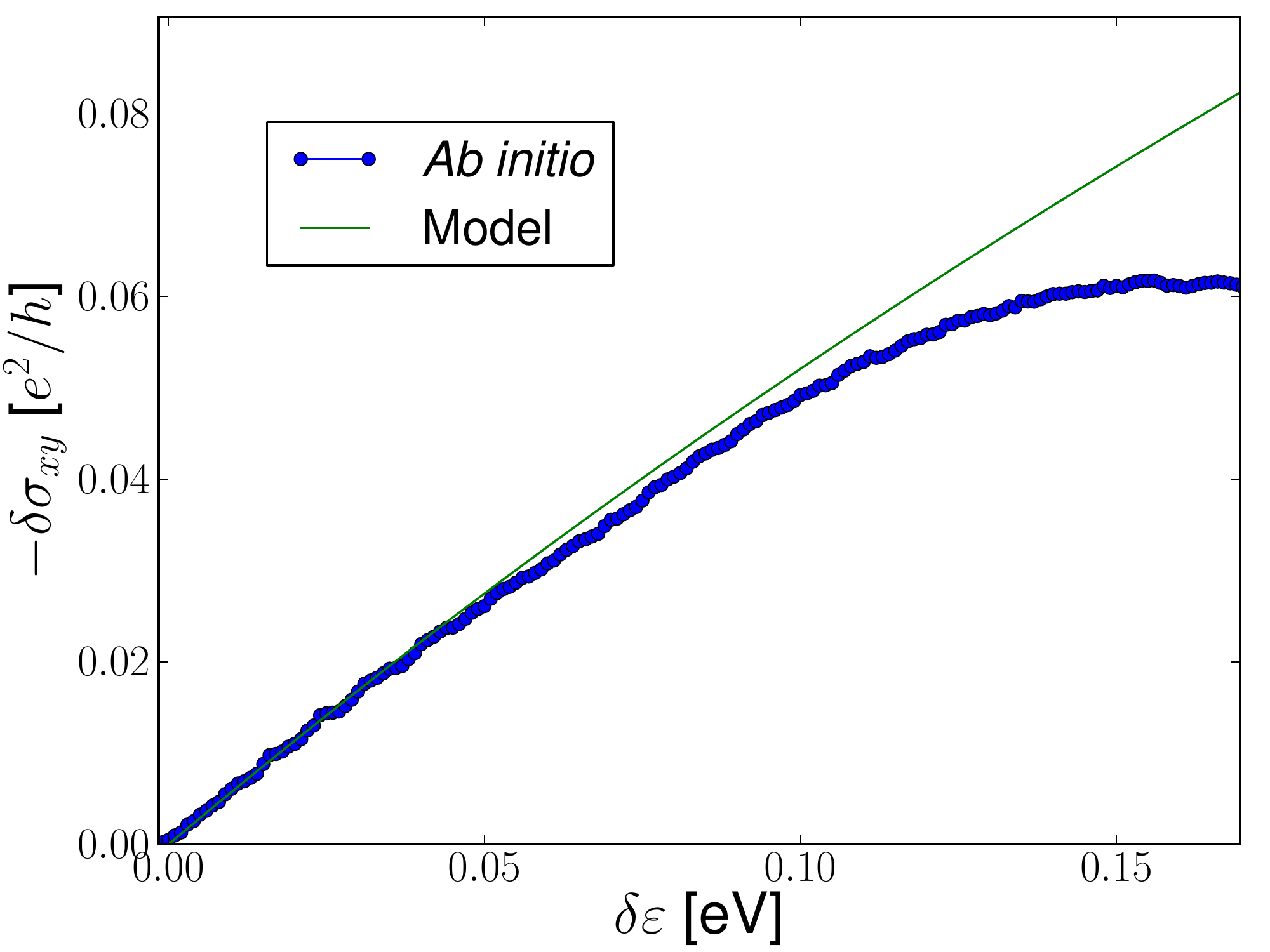}
    \caption{(Color online). Intrinsic photoinduced
      anomalous Hall conductivity of a single
      layer of MoS$_2$ without
        gating ($\ef=0$), plotted as a function of the Fermi-level shift $\de$ in the $K$
        valley. The thin green line corresponds to Eq.~\eqref{ahc_linear} for the massive Dirac
        model with the parameters given in the text, and the blue
        dotted line was obtained by evaluating
        \eqs{vhc-KL}{photo-ahc} from
        first-principles.}
\label{fig:pristine_vhc}
\end{figure}

In Fig.~\ref{fig:pristine_vhc} we show the intrinsic
photoinduced AHC $\delta\sigma_{xy}^{0}$ from a single
spin channel
%
%\ism{In my opinion it is strange to show ab initio results for a
%  physical quantity only taking into account a single spin
%  channel. Presumably what you did was to divide the ab initio number
%  by two, to compare with the Dirac model. A better option would be to
%  do the opposite: multiply the result of the model calculation by
%  two. As a bonus, you would not have to keep mentioning ``single spin
%  channel.'' at several points in the manuscript.}
%
as a function of the single-valley energy shift $\de$,
for $\varepsilon_F=0$. It should be noted that due to the spin-orbit split 
valence bands in MoS$_2$, it is possible to selectively excite a single spin channel
by using an optical frequency tuned to the transition between the topmost
valence band and the conduction band.\cite{xiao_valley, mak_valley}
%
%\ism{I am confused about the following: you say here that the
%  calculations were done for $\ef=0$. But in
%  Sec.~\ref{sec:dirac-model} you only give expressions for the
%  photoinduced AHC of the Dirac model with {\it nonzero} $\ef$, in
%  fact for $\ef>\Delta$ where $\Delta=0.86$~eV.  Isn't the slope of
%  the green line in Fig.~\ref{fig:pristine_vhc} equal to the prefactor
%  $e^2\Delta/(h\ef^2)$ of the middle expression in \eqref{ahc_linear}?
%  The quantity $\ef$ appears in the denominator, so surely it was not
%  evaluated for $\ef=0$. I don't quite understand what is being done,
%  I feel it needs to be explained a bit more carefully.}
%
 The \textit{ab initio} calculations are seen to
agree very well with the model results at the bottom of the
valley. When the energy shift approaches $\sim0.1$ eV the
secondary valley located between $\Gamma$ and $K$ starts to
contribute, and the model results become unreliable.

The measurements of the valley Hall effect reported in
Ref. \onlinecite{mak_valley} involved photoexcitation of states in a
single valley. In that work, the carrier density was estimated
from photoconductivity measurements, and the photoinduced AHC was
displayed as a function of carrier density. The typical density
  of photoexcited carriers reported in Ref.~\onlinecite{mak_valley}
  is of the order of $10^{-11}$
cm$^{-2}$,
%
%  \ism{You had written $10^{-11}$ cm$^{-2}$\cite{mak_valley}, where
%    the 2 in the exponent got combined with the 10 from the reference
%    number.}
%
  which corresponds to an energy shift
    $\de\sim 1$~meV; this is far below the point
  where the linear model \eqref{ahc_linear} breaks down. 
	
	%We note that in Ref. \onlinecite{mak_valley} Fig. 3, the photoinduced AHC was
  %measured as a function of estimated carrier imbalance for different
  %gate voltages. They found a linear dependence with the slope
  %depending on the gate voltage, which is exactly what is contained in
  %Eq. \eqref{ahc_linear}. However, since a certain gate voltage cannot
  %directly be converted into a Fermi level shift, a direct comparison
  %is not possible.

\section{Disordered MoS$_2$}\label{disordered}

Experimental as well as theoretical studies has recently demonstrated
that sulfur vacancies are the dominant source of disorder in
MoS$_2$.\cite{zhou_defects, liu_defects, ma_defects, zhou_defects1,
  wei_defects, asl_defects} In the following we will thus exclusively
focus on the sulfur vacancies and calculate how they affect the VHC. 
The results will be analyzed in terms of an
impurity averaged unfolded Berry curvature, which will be defined
below. However, before we delve into the \textit{ab initio}
calculations we will briefly review the theoretical results for the
AHC in a massive Dirac model with impurity scattering.

\subsection{Massive Dirac model}

To obtain a model result for the VHC of disordered MoS$_2$, we use the
extrinsic contribution to the AHC of the massive Dirac Hamiltonian
\eqref{H_dirac}, which has been calculated in the limit of weak and
dilute scattering.\cite{sinitsyn_link} The impurity potential for a
given configuration is
$V(\mathbf{r})=\sum_iV_i\delta(\mathbf{r}-\mathbf{R_i})$, where
$\mathbf{R_i}$ are impurity sites. The result for the intrinsic
(0), side jump (SJ) and skew scattering (SS)
contributions to the VHC of the $K$~valley is then
%
%\ism{Regarding these three equations:\\
%  $\bullet$ Eq.~\eqref{intrinsic} is equivalent to \eq{vhc-dirac}, but
%  this is not mentioned. Moreover, a different symbol is used on the
%  left-hand-side (LHS). One possible solution would be to write on the
%  LHS of both \eq{vhc-dirac} and Eq.~\eqref{intrinsic}
%  $\sigma_{xy}^{K,{\rm KL}}$, where KL stands for
%  ``Karplus-Luttinger,'' which is often done in the literature. Then
%  we should
%  also write $\sigma_{xy}^{\rm KL}$ in \eq{ahc}, and cite the Karplus-Luttinger paper there.\\
%  $\bullet$ Notation should be changed from $\sigma_{VH}$ to
%  $\sigma_{xy}^K$.  These expressions are written for the $K$ valley,
%  right? We should state how the corresponding terms for the $K'$
%  valley are related to the ones we give. Do the SJ and SS terms flip
%  sign, like the intrinsic term?}
%
\begin{align}\label{intrinsic}
\sigma^{0,K}_{xy} = 
\frac{e^2\Delta}{2h\sqrt{(v\hbar k_F)^2+\Delta^2}},
\end{align}
\begin{align}\label{SJ}
\sigma^{SJ,K}_{xy} = 
\frac{e^2\Delta}{2h\sqrt{(v\hbar k_F)^2+\Delta^2}}\bigg[&\frac{4(v\hbar k_F)^2}{4\Delta^2+(v\hbar k_F)^2}\\
&+\frac{3(v\hbar k_F)^4}{(4\Delta^2+(v\hbar k_F)^2)^2}\bigg],\notag
\end{align}
\begin{align}\label{SS}
\sigma^{SS,K}_{xy} = 
\frac{e^2\langle V_i^3\rangle_c\Delta}{hx\langle V_i^2\rangle_c^2}\frac{(v\hbar k_F)^4}{(4\Delta^2+(v\hbar k_F)^2)^2}.
\end{align}
%eq:vhc-dirac
Equation~\eqref{intrinsic} is just \eq{vhc-dirac} recast in a
  different form. As mentioned in the Introduction, scattering
  contributions which are independent of the impurity
  concentration~$x$ 
%
%  \ism{How is $x$ defined exactly? Number of impurities per primitive
%    cell, or per site? (There are two sites, A-sublattice and B
%    -sublattice, in each cell.)}
%
are classified as side-jump, and those which scale
inversely with~$x$ are classified as skew-scattering.  
In particular, the second term in Eq.~\eqref{SJ} originates from a fourth-order
expansion of the scattering rate, and could thus be regarded as a
skew-scattering contribution. Furthermore, the side-jump contribution
contains contributions that cannot directly by ascribed to a
coordinate shift.\cite{sinitsyn_link} 
%
%\ism{This sentence seems to be saying basically the same thing as the
%  previous one. Or are you now referring to the first term in
%  Eq.~\eqref{SJ}?}
%

\begin{figure}[tb]
    \includegraphics[width=8.0 cm]{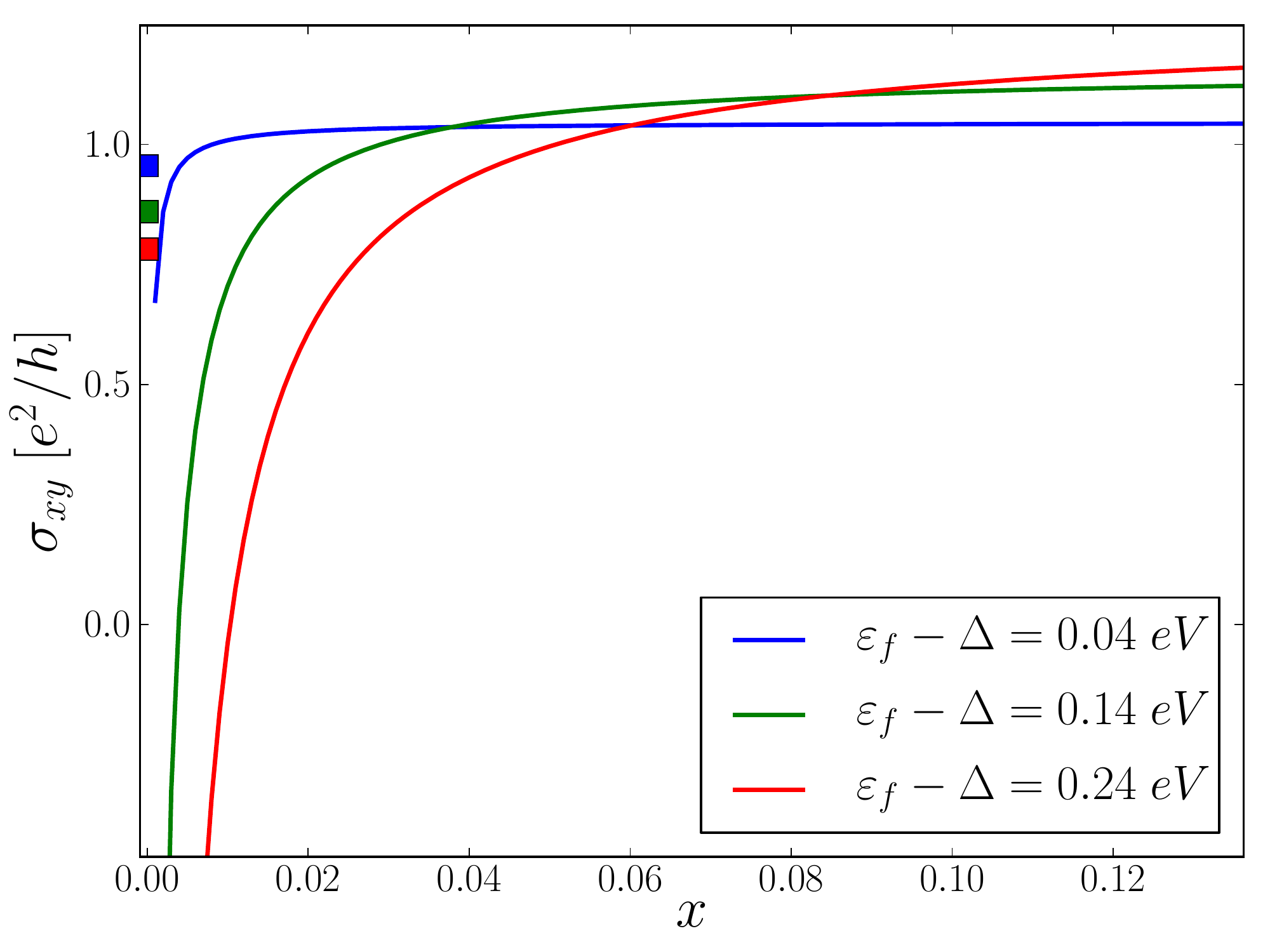}
    \caption{(Color online). Valley Hall conductivity of
      the disordered massive Dirac model,
      Eqs.~\eqref{intrinsic}-\eqref{SS}, evaluated as a function of impurity
      concentration~$x$ for three different gating levels, with
      $\langle V_i^3\rangle_c/\langle V_i^2\rangle_c^2=-0.4$~eV and
      $\Delta=0.86$~eV. The squares at $x=0$ correspond to the intrinsic contribution
        of Eq.~\ref{intrinsic}.}
\label{fig:hall_model}
\end{figure}

In Fig. \ref{fig:hall_model} we show the VHC as a
function of impurity concentration at different Fermi level shifts
calculated within the model with $\Delta=0.86$ eV and $\langle
V_i^3\rangle_c/\langle V_i^2\rangle_c^2=-0.4$ eV$^{-1}$. The
VHC converges to the side jump at large impurity
concentrations, but one should keep in mind that the model results are
derived under the assumption of dilute disorder and weak
scattering. The side jump correction to the intrinsic part becomes
larger for higher density of states at the Fermi level and is
neglectable at very low carrier concentrations. Furthermore, at low
carrier concentrations, the skew scattering only becomes significant
at very low impurity concentrations. It thus appears that for low
carrier concentrations, the intrinsic contribution gives a good
account of the AHC - even at rather low impurity
concentrations.

\subsection{Unfolded band structure and Berry curvature}

In pristine systems, the Berry curvature provides a useful
$k$-resolved measure of the Hall conductivity. For example, from
Fig. \ref{fig:pristine_band_curv} it is clear that states near the
valleys at $K$ and $K'$ 
%
%\ism{In this section the symbol $K$ is used with two different
%  meanings, first as a valley label in the NBZ, then as a generic
%  point in the SBZ.  The conflict can be resolved by writting the
%  latter more correctly as ${\bf K}$. I have tried to do that below,
%  but please check.}
%
have the largest potential for contributing to
the VHC. However, if we would like to know how a given
distribution of impurities affects the VHC, this
picture immediately breaks down since the pristine Brillouin zone is
no longer relevant due to broken translational symmetry. On the other
hand, if a given impurity distribution is represented in a supercell,
the VHC will still be given as a $k$-space
integral of the Berry curvature, but now the domain will be the
Brillouin zone corresponding to the supercell (SBZ), which is not
directly comparable to the normal Brillouin zone (NBZ). Nevertheless,
we can expand the supercell states in terms of states in the pristine
system and thus unfold the supercell curvature to the pristine
Brillouin zone. For a general band quantity $a_N(\K)$ defined in SBZ we
can thus define the unfolded quantity $a^{(u)}_n(\k)$ in the NBZ as
\begin{align}\label{eq:unfold_general}
a^{(u)}_n(\k)=\sum_N|\langle N\K|n\k\rangle|^2a_N(\K),
\end{align}
where $\K$ is the SBZ crystal momentum, which is related to $k$ by
translation of a supercell reciprocal lattice vector. Brillouin zone
integrals can be written in terms of the unfolded function since
\begin{align}\label{unfolding}
A&=\int_{\rm SBZ}d\K\sum_Na_N(\K)\notag\\
&=\int_{\rm SBZ}d\K\int_{\rm NBZ}d\k\sum_{nN}|\langle N\K|n\k\rangle|^2a_N(\K)\notag\\
&=\int_{\rm NBZ}d\k\sum_na^{(u)}_n(\k)\,,
\end{align}
where a complete set of NBZ states were inserted in the second line
and it was used that $\langle N\K|n\k\rangle$ is only non-vanishing if
$\k$ downfolds to $\K$.

\begin{figure}[tb]
    \includegraphics[width=8.0 cm]{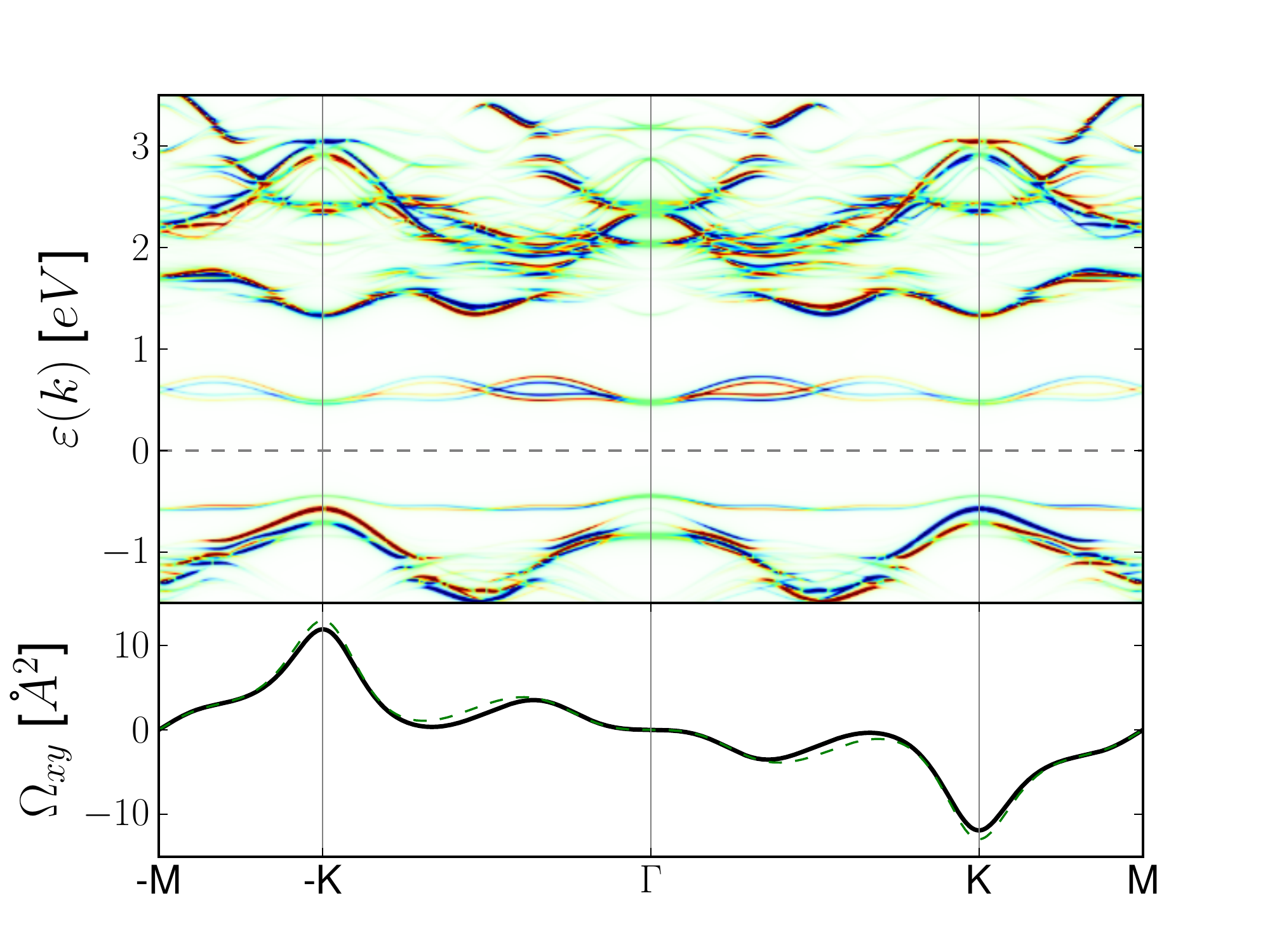}
    \caption{(Color online). Unfolded spectral function
      (top) and Berry curvature (bottom) of a
      $3\times3$ MoS$_2$ supercell with a single
      sulfur vacancy. The Fermi level $\ef=0$ is in the gap and
      the Berry curvature is largely unaffected by the presence
        of the impurity (the Berry curvature for pristine \mos\,,
        taken from Fig.~\ref{fig:pristine_band_curv}, is drawn as a
        dashed green line).}
\label{fig:MoS2_33}
\end{figure}

The method of unfolding $k$-space quantities has previously been
applied to band structures of disordered systems\cite{ku_unfolding,
  berlijn_effective} and more recently to unfolding the Berry
curvature.\cite{bianco_unfolding} In the case of band structures the
object of interest is the spectral function
\begin{align}\label{unfold_bands}
A^{(u)}(\omega,\k)=\sum_{nN}|\langle N\K|n\k\rangle|^2\delta(\omega-\varepsilon_{N\K}).
\end{align}
The Berry curvature is somewhat more complicated since, a naive application
of \eq{unfold_general} leading to
%
%\ism{I would prefer removing the rest of this Appendix, and simple
%  referring to the original paper. I don't think this is equivalent to
%  what is in our paper, and the expression in our paper is in fact
%  what you actually implemented. This expression that you show was of
%  course our first idea when we started the project, but then we
%  realized it didn't work.}
%
\begin{align}\label{unfold_curvature}
\Omega^{(u)}(\k)=\sum_{nN}|\langle N\K|n\k\rangle|^2f_{N\K}\Omega_N(\K),
\end{align}
becomes gauge dependent. In
Ref. \onlinecite{bianco_unfolding} Eq. \eqref{unfold_curvature} it was shown h
how to generalize Eq. \eqref{eq:unfold_general} to obtain an explicitly gauge
invariant quantity. In the present work we have applied the gauge invariant 
expression for all calculations.

\subsection{Sampling impurity configurations}

\begin{figure}[tb]
    \includegraphics[width=8.0 cm]{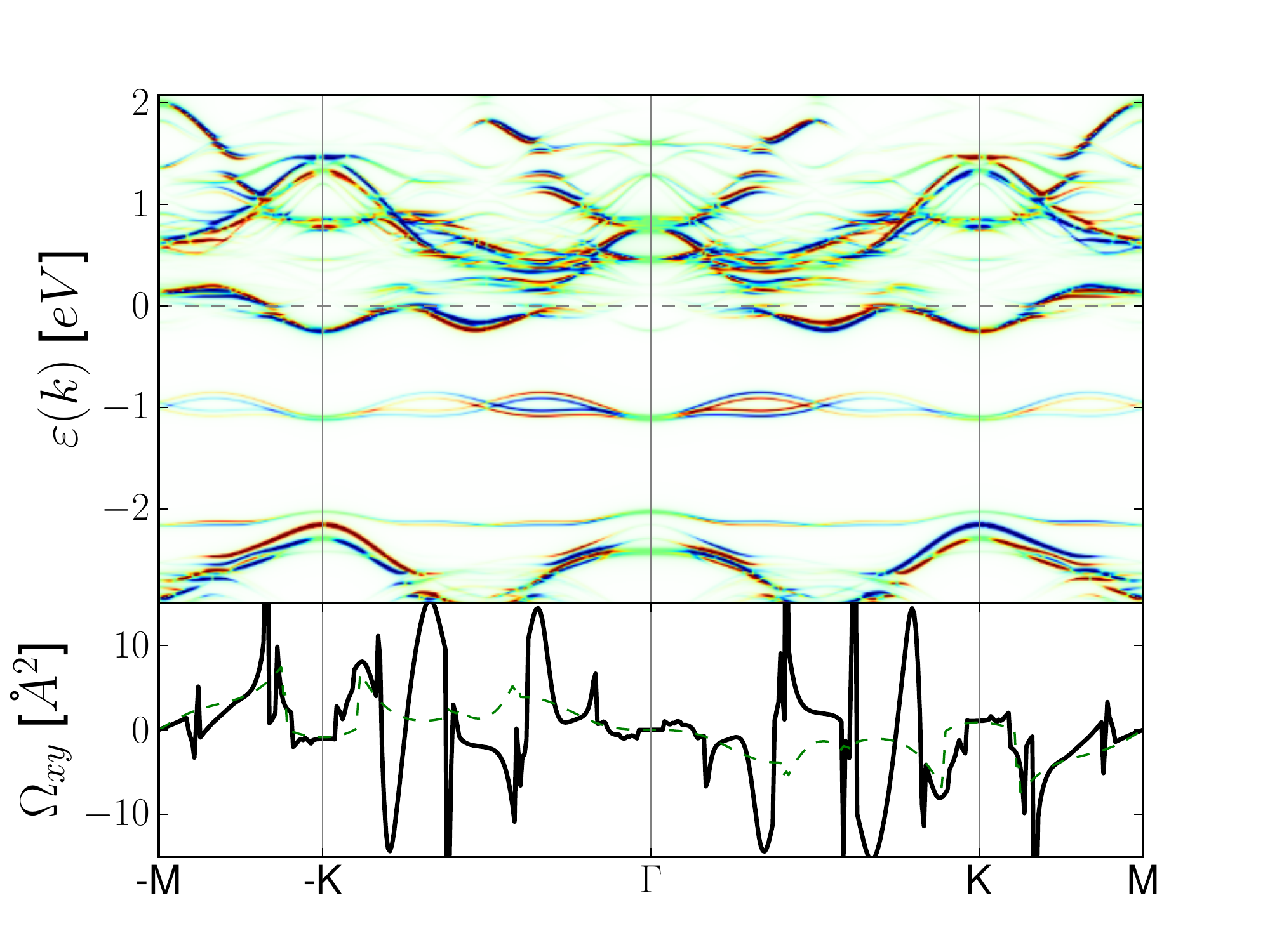}
    \caption{(Color online). Same as Fig.~\ref{fig:MoS2_33},
        except that the Fermi level has been shifted to the
      conduction band, where the impurities induce large modifications
      of the Berry curvature.}
\label{fig:MoS2_33_shifted}
\end{figure}

In Fig. \ref{fig:MoS2_33} we show the unfolded spectral function and
Berry curvature of a periodic structure obtained as $3\times3$ MoS$_2$
unit cell with a single sulfur vacancy. The vacancy is seen to
introduce both occupied and unoccupied states in the gap, but the
Berry curvature is largely unaffected by such rather localized
states. This system is a example of a $x=1/18=0.56$ impurity
concentration, but is not necessarily representative of this disorder
concentration in general. As it turns out the Berry curvature is
largely insensitive to the impurity configuration as long as the Fermi
level is in the gap. However, the situation changes dramatically when
the Fermi level is shifted to the conduction band. This situation is
shown in Fig. \ref{fig:MoS2_33_shifted}. The blurred features in the
unfolded spectral function is associated with scattering states and
are accompanied by spiky feature in the Berry curvatures. From the
spectral representation of the Berry curvature Eq. \eqref{berry_tb},
it is clear that such features arise whenever occupied and unoccupied
states come close to the Fermi level.

In experiments with MoS$_2$ transistors the Fermi level is typically
controlled with a gate voltage.\cite{mak_valley} Furthermore,
exfoliated MoS$_2$ usually exhibits an intrinsic $n$-doping, which is
attributed to Re impurities.\cite{komsa} For these reasons, we have
chosen to pin the Fermi level to the conduction band in the
following. This will facilitate the comparison with experiments as
well as model calculations.

\begin{figure}[tb]
    \includegraphics[width=8.0 cm]{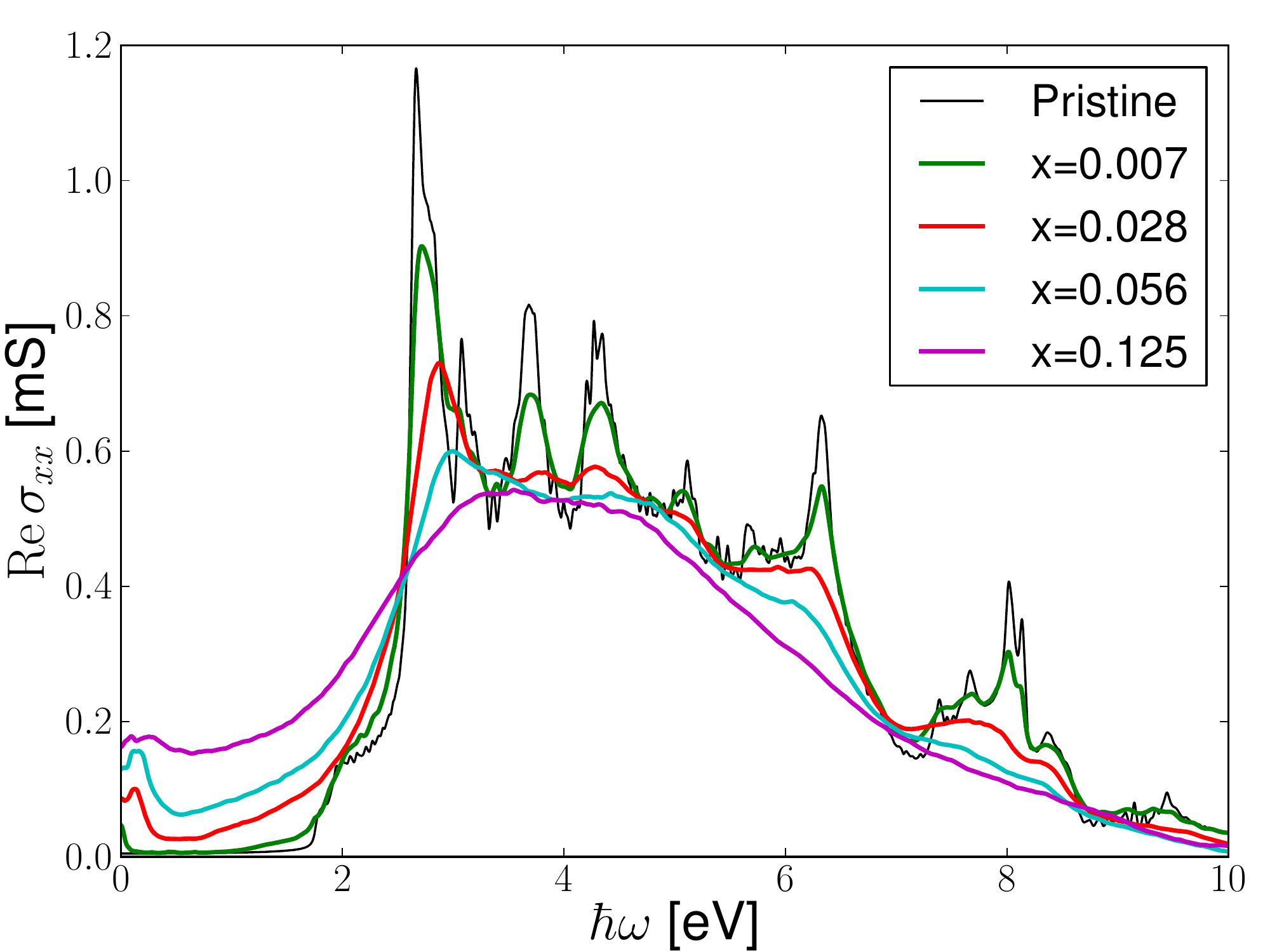}
    \caption{(Color online). Configuration-averaged optical absorption
      spectrum of MoS$_2$ at various impurity
      concentration.}
\label{fig:MoS2_opt}
\end{figure}

In order to perform a faithful calculation of the conductivity at a
given impurity concentration, one then has to average over a large
number of impurity configurations. Such a procedure requires a high
number of simulations in large unit cells and is not feasible with
standard \textit{ab initio} methods. To proceed we construct an
effective Hamiltonian based on \textit{ab initio} DFT calculations and
Wannier functions. Given an impurity concentration, we thus randomly
generate an impurity configuration and construct the corresponding
Hamiltonian as described in Appendix \ref{effective}. For a given
vacancy concentration, we perform calculations for $\sim1000$ systems
with randomly drawn impurity configurations in a $6\times12$ supercell. 
The smallest impurity concentration considered was $x=1/216$,
where we needed a larger supercell of $12\times18$. As a first check
of the method, we calculate the optical conductivity of MoS$_2$ at
various S vacancy concentrations. The results are shown in
Fig. \ref{fig:MoS2_opt} and as expected the impurities simply
introduce a broadening of the spectra. We should note that 
the Wannier functions used to construct the tight-binding Hamiltonian
was obtained by disentangling bands up to 3.0 eV above the conduction band 
minimum and  the present calculation is thus only expected to be reliable up
to 3.0 eV above the pristine absorption edge.

\subsection{Valley Hall conductivity}

\begin{figure*}[tb]
    \includegraphics[width=8.7 cm]{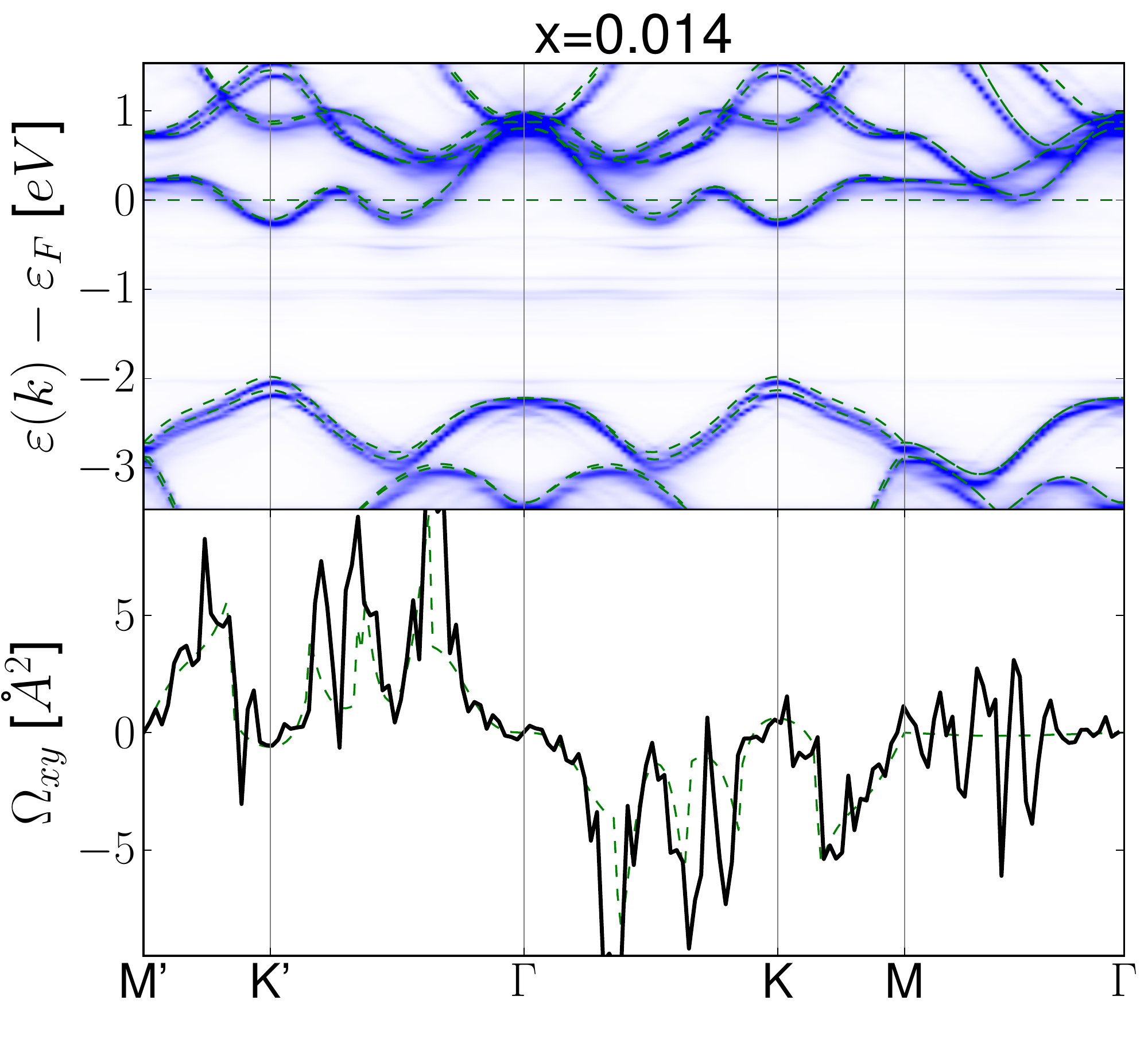}
    \includegraphics[width=8.7 cm]{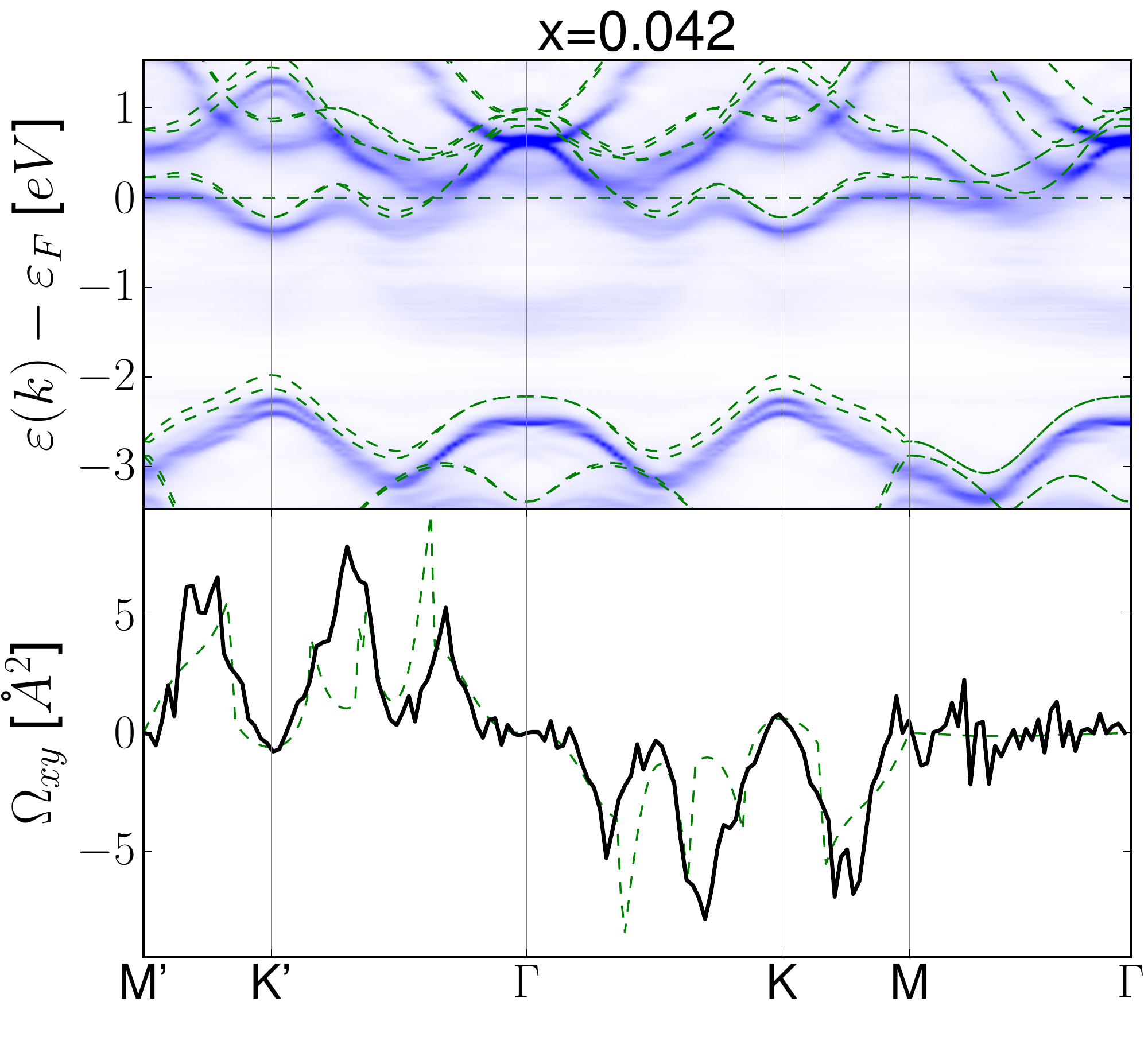}\\
    \includegraphics[width=8.7 cm]{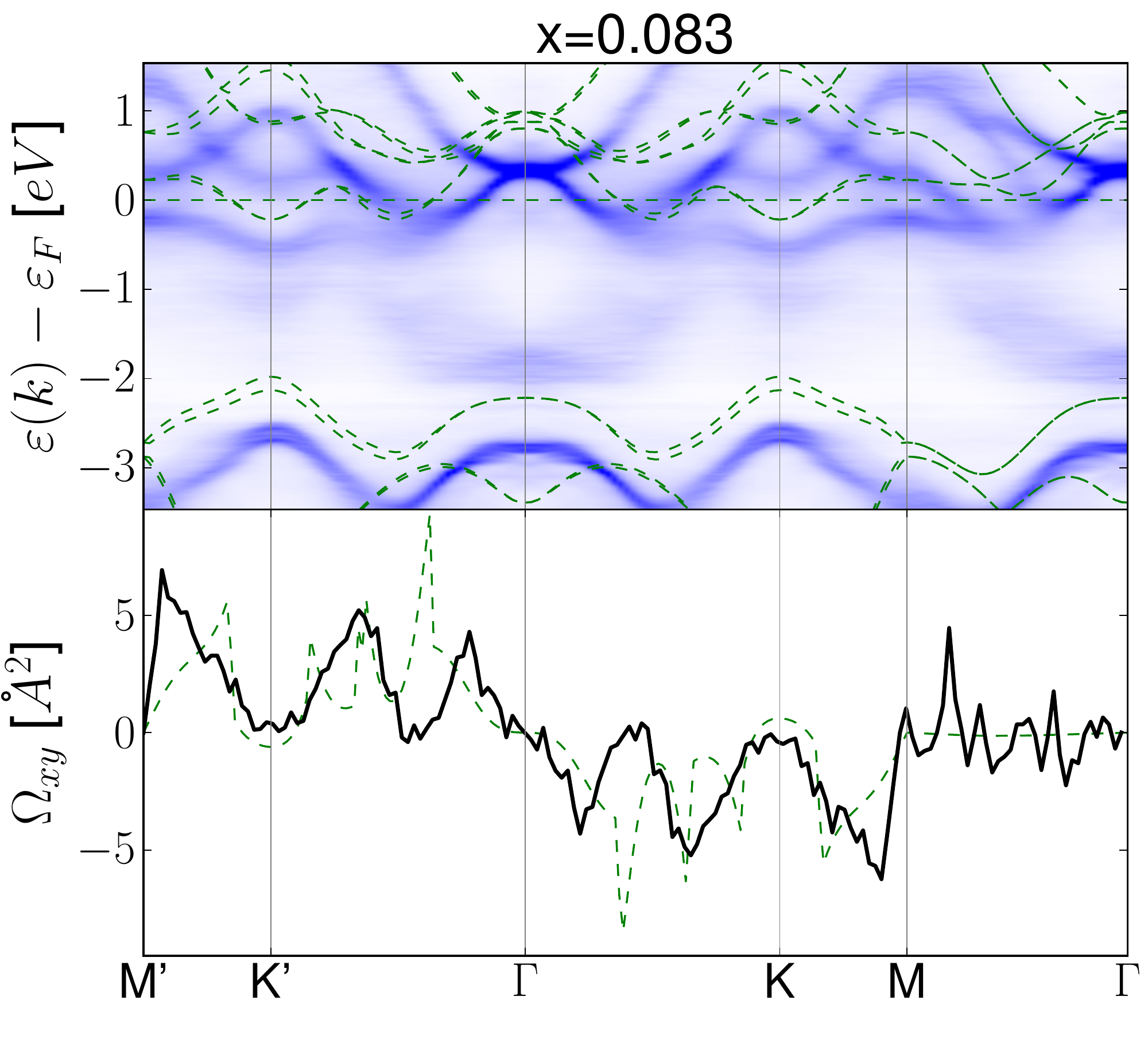}
    \includegraphics[width=8.7 cm]{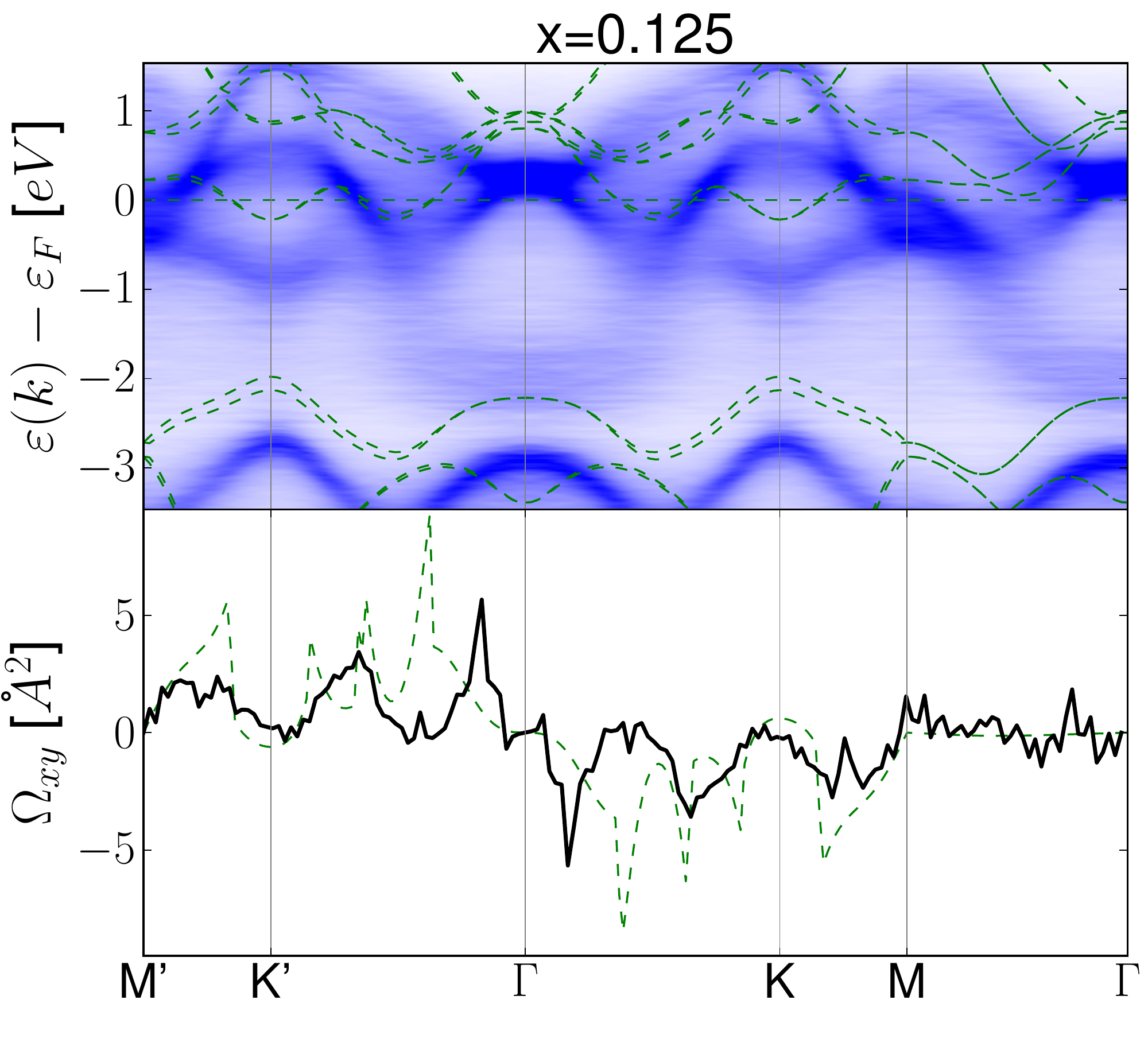}
    \caption{(Color online). Impurity averaged unfolded spectral
      function and Berry curvature of disordered MoS$_2$ with the
      Fermi level fixed in the conduction band. The dashed green lines
      shows the result for pristine MoS$_2$.}
\label{fig:MoS2_dis_curv}
\end{figure*}

The lowest impurity concentration we have considered is
$x=0.005$. Here the configurational averaged unfolded Berry curvature
becomes very spiky and is not particularly informative. In
Fig. \ref{fig:MoS2_dis_curv} we show the configurational averaged
unfolded Berry curvature for 4 different intermediate impurity
configurations for $\varepsilon_F-\Delta=0.24$ and
$\varepsilon_F-\Delta=0.24$. In general we observe that the peaks in
the Berry curvature becomes enhanced and broadened, while the
curvature retains its qualitative features. This tendency is
maintained over a range of impurity concentrations ($x=0.02$ -
$x=0.1$) and gives rise to a impurity independent increase in VHC. 
We will identify this with the side jump corrected
VHC and the Berry curvatures in
Fig. \ref{fig:MoS2_dis_curv} thus comprises a measure of the $k$-space
resolved side jump scattering. From the unfolded spectral function it
is clear that the sulfur vacancies has the effect of lowering the
overall potential, such that the conduction bands are lowered with
respect to the fixed Fermi level at high impurity
concentrations. However, for intermediate dopings the exact position
of the Fermi level does not have a large effect on the VHC. 
In fact, raising the Fermi level with respect to the
conduction band tends to lower the VHC, since
more curvature from the conduction band will be included and this has
the opposite sign of the dominating contribution from the valence
bands. At larger impurity concentrations ($x=0.125$) the Berry
curvature becomes more "smeared" and will eventually average to
zero. At this point the system is so strongly perturbed that it cannot
be analyzed in terms of scattering events.

\begin{figure}[t]
    \includegraphics[width=8.0 cm]{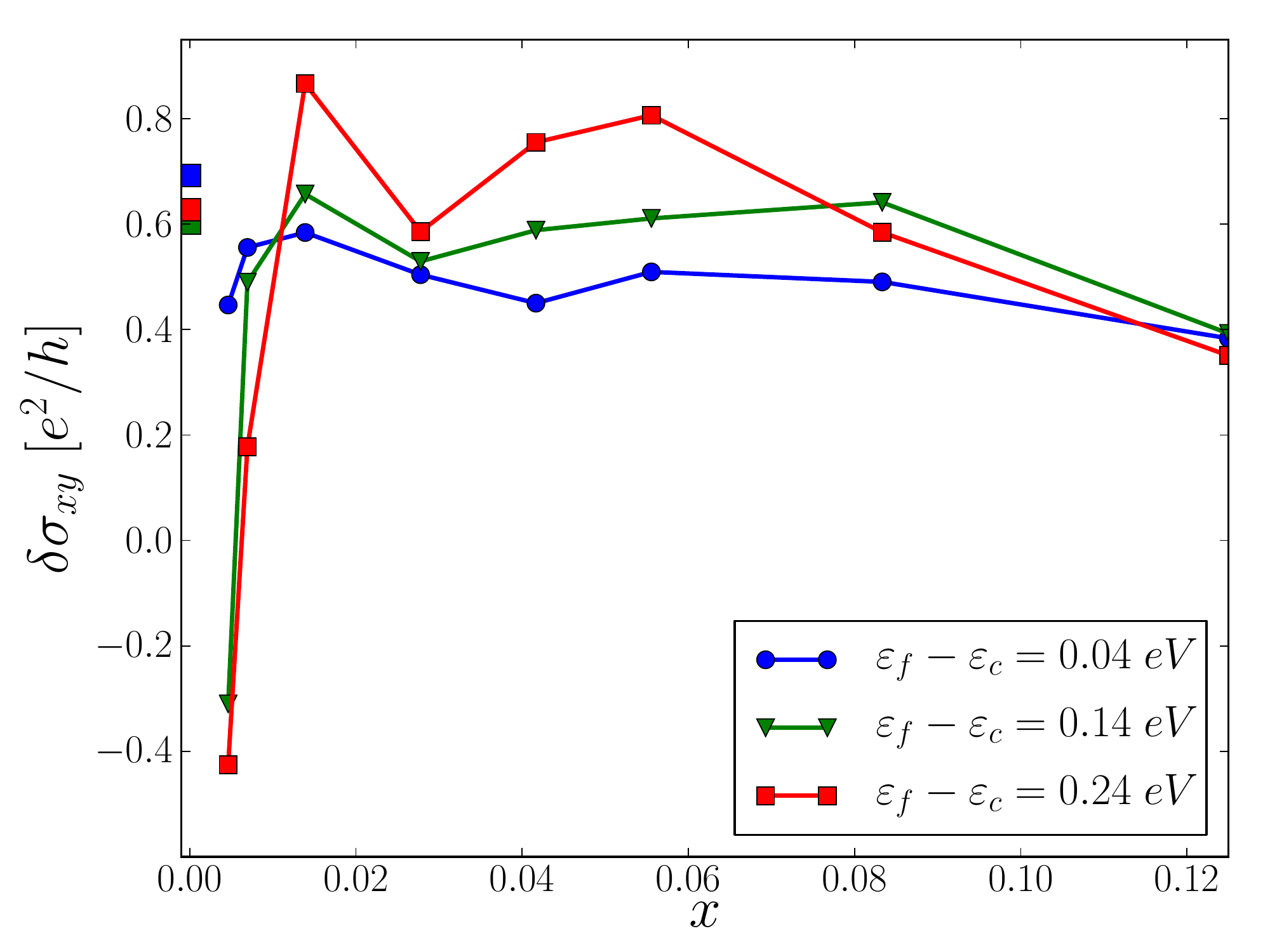}
    \caption{(Color online). Calculated VHC as a
      function of impurity concentration $x$. We show the result for
      three different positions of the Fermi level with respect to the
      conduction band minimum $\varepsilon_c$.}
\label{fig:hall_impurity}
\end{figure}

In Fig. \ref{fig:hall_impurity} we show the VHC as a function of
impurity concentration at different Fermi level shifts. At low
impurity concentrations we observe a divergent behavior of the
VHC. Comparing with Eqs.  \eqref{intrinsic}-\eqref{SJ} we identify
this with skew scattering processes. In terms of the unfolded
curvature, this is the "spiky" regime, where the averaged curvature
largely looses the qualitative features of the pristine system. We
note that as the impurity concentration $x$ is decreased, it becomes
progressively harder to converge the results. This is due to the fact
that the standard deviation for a given supercell
size increases when $x$ becomes small, while at the same time we need
very large supercells in order to perform
calculations for small $x$.

At intermediate impurity concentrations the VHC reaches a flattened
plateau, which we identify as the side jump regime. In
  this regime skew scattering is insignificant and the pristine
result receives a small correction which is nearly independent of
impurity concentration. Comparing with Fig. \ref{fig:hall_model}, we
see that there is good qualitative agreement between the model and the
\textit{ab initio} results. A major difference is the fact that the
\textit{ab initio} VHC decreases at high impurity concentrations,
whereas the model converges towards the side jump result. The reason 
is that the model results were
derived under the assumption of dilute impurity concentration, and
cannot be applied in this regime. Moreover, at large Fermi level
shifts the deviations from the Dirac model begins to be important. In
the present system the most important effect is the contribution from
the secondary conduction band minimum between $\Gamma$ and $K$.

When the Fermi level comes close to the conduction band minimum, two
crucial features can be observed from Figs. \ref{fig:hall_model} and
\ref{fig:hall_impurity}. First, the side jump correction becomes small
such that the intrinsic VHC becomes a good estimate at
intermediate impurity concentrations. Second, the critical impurity
concentrations, where skew scattering starts to dominate, becomes
rather small. For $\varepsilon_F-\varepsilon_c=0.04$ this
concentration is $x\sim0.005$ and will be even smaller when the Fermi
level moves closer to the band edge.

\subsection{Photoinduced anomalous Hall conductivity}\label{sec:photo_ind}
As mentioned previously, the experimentally measured quantity is the
photoinduced AHC, which we express as the sum of the two VHCs.  For
small carrier imbalance this will be
$\delta\sigma_{xy}(\varepsilon_F,\delta\varepsilon)=d\sigma_{VH}(\varepsilon)/d\varepsilon|_{\varepsilon=\varepsilon_F}\delta\varepsilon$,
where $\delta\varepsilon$ is the shift in Fermi level corresponding to
the charge imbalance. A first principles evaluation of this
expression including impurities is made difficult by the fluctuations
involved in the configurational averaging procedure. However, the good
agreement between the \textit{ab initio} calculations of the VHC
(Fig. \ref{fig:hall_impurity}) with the massive Dirac model
(Fig. \ref{fig:hall_model}), suggests that we can obtain a reliable
estimate of the photoinduced AHC from the Dirac model. For the skew
scattering contribution we take $\langle V_i^3\rangle_c/\langle
V_i^2\rangle_c^2=-0.4$ eV as in Fig. \ref{fig:hall_model}. Note that
this value, and in particular the sign, is a non-trivial \textit{ab
  initio} result obtained from the VHC calculations.

\begin{figure}[t]
    \includegraphics[width=7.0 cm]{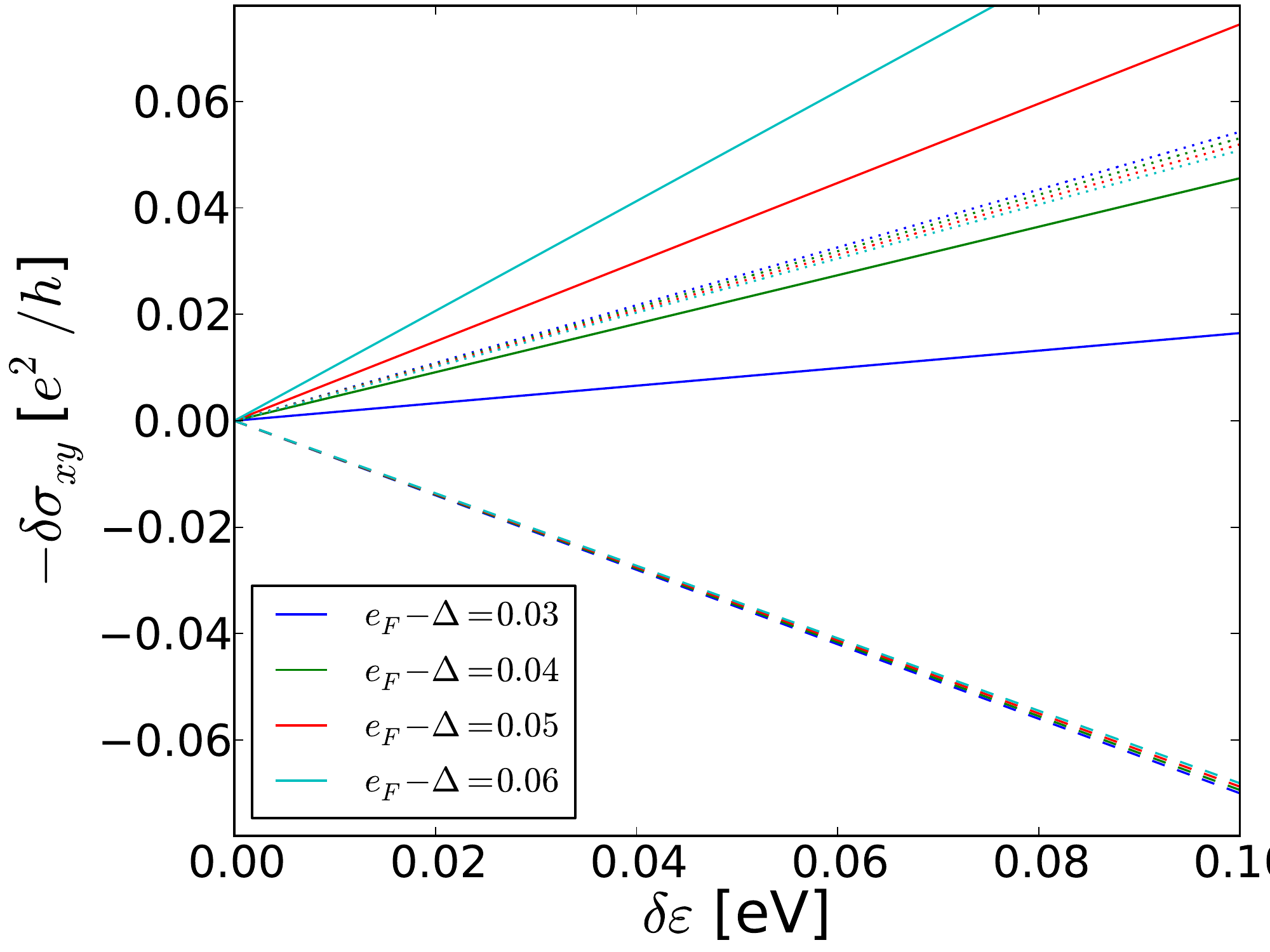}
    \caption{(Color online). Photoinduced AHC as a function of Fermi
      level shift $\delta\varepsilon$ between the valleys. States in
      the K valley are assumed excited corresponding to application of
      a left-handed polarized field. The dotted lines are the intrinsic
      $\delta\sigma^0_{xy}$, the dashed lines are $\delta\sigma^0_{xy}+\delta\sigma^{SJ}_{xy}$,
      and the solid lines are
      $\delta\sigma^0_{xy}+\delta\sigma^{SJ}_{xy}+\delta\sigma^{SS}_{xy}$. Note the large
      dependence on Fermi level when skew-scattering is included.}
% \label{fig:hall_impurity}
    \label{fig:vhc_scat}
\end{figure}

In Fig.~\ref{fig:vhc_scat}
%
%\ism{Is this the figure you wanted to refer to? Because this figure
%  has a duplicate label {\tt fig:hall\_impurity}.}
%
we show the photoinduced AHC calculated from
Eqs. \eqref{intrinsic}-\eqref{SS} with $x=0.01$. First of all
it should be noted that the inclusion of side jump scattering gives a
result, which is very close to the intrinsic contribution, except that
the sign has been changed. This was already noted in
Ref. \onlinecite{mak_valley} For the VHC, the side jump only
contributes a minor part to the scattering independent VHC. However,
for the derivative, which is the relevant quantity for the
photoinduced AHC, the side jump plays a dominant role. At small
impurity concentrations, the skew scattering contribution will
dominate. The skew scattering part of the VHC scales as $1/x$ and so
does its contribution to the photoinduced AHC. Interestingly, the skew
scattering contribution also has significant dependence on the
position of the Fermi level and could explain the large dependence on
gate voltage observed experimentally.\cite{mak_valley}

\section{Conclusions}
\label{sec:conclusions}

\textit{A priori} it is not clear that the intrinsic VHC provides a
good descriptor for the VHC in a MoS$_2$ transistor setup, since the
Hall conductivity is expected to diverge in the clean limit as a
consequence of skew scattering.\cite{mak_valley} Here we have used
first principles calculations to investigate the effect of sulfur
vacancies at different impurity concentrations. The influence of
disorder was analyzed in $k$-space in terms of the unfolded Berry
curvature and we have shown that the side jump regime appears as a
concentration independent enhancement of the Berry curvature. The skew
scattering introduces divergences in the Berry curvature and the
unfolded Berry curvature becomes spiky and irrelevant. Nevertheless,
we were able to converge the VHC calculations in the skew scattering
regime, and recover the expected divergent behavior. The \textit{ab
  initio} calculations show qualitative agreement with model
calculations based on a massive Dirac Hamiltonian and allow us to
extract a non-trivial value of the skew scattering potential $\langle
V_i^3\rangle_c/\langle V_i^2\rangle_c^2=-0.4$ eV.  The calculations
allow us to estimate a critical impurity concentration $x$ where skew
scattering starts to dominate. For $\varepsilon_F=40$ meV (with
respect to the conduction band minimum) we find that $x\sim0.005$ and
below this point the intrinsic VHC becomes a poor descriptor.

A comparison with experiments\cite{mak_valley} indicates that indeed,
the intrinsic VHC cannot be applied as a descriptor for the
photoinduced valley Hall conductivity. As previously noted, the side
jump contribution changes the sign of photoinduced AHC and we have
shown that the skew scattering contribution is a likely explanation
for the large gate dependence observed
experimentally.\cite{mak_valley} However, a reliable estimate of the
skew scattering contribution requires knowledge of the impurity
concentrations in the samples investigated, which is not presently
available. It would be very interesting to perform measurements of the
photoinduced AHC on MoS$_2$ samples with different impurity
concentrations in order to unravel the roles played by side jump and
skew scattering.

We have implicitly assumed a low temperature regime and therefore not
discussed the effect of phonons. The effect of phonons on the
longitudinal mobility in MoS$_2$ has been analyzed
thoroughly,\cite{kristen} but the influence on the transverse
conductivity has so far not been considered. Furthermore, monolayer
MoS$_2$ have been shown to exhibit strong excitonic effects
\cite{huser_MoS2, louie_MoS2} due to poor screening in two-dimensional
materials. The charge imbalance utilized in the experimental
realization originates from optically generated electron-hole pairs,
and if the Fermi level is close to the conduction band edge these
effects could potentially severely limit the carrier mobility. We will
leave these issues for future studies.

\section{Acknowledgement}

This work was supported by the Danish Council for Independent Research,
Sapere Aude Program,
and by grants No.~MAT2012-33720 from the Ministerio de
  Econom\'ia y Competitividad (Spain) and No.~CIG-303602 from the
  European Commission.

\appendix

\section{Calculational details}
The calculations in the present work was performed with the tight
binding method using parameters obtained from \textit{ab initio}
density functional calculations and projected Wannier functions.

\subsection{\textit{Ab initio} calculations and construction of
    Wannier orbitals}
\label{sec:abinitio}

The \textit{ab initio} density functional theory calculations were
performed with the {\tt pwscf} code from the {\tt Quantum
  Espresso} package,\cite{quantum_espresso} using the PBE functional.
%
%\ism{Reference needed.}
%
 Norm-conserving pseudopotentials
were used,
%
%\ism{Provide a reference?}
%
and the calculations were carried out with a planewave cutoff of
100~Ry. The lattice parameter of MoS$_2$ was set to the experimental 
lattice constant of 3.16~{\AA},
%
%\ism{Is this the experimental lattice constant, or the optimized
%  theoretical one?}
%
  and 12~{\AA} was used to separate the periodically-repeated
  images. All calculations were performed in a non-collinear
    spin framework, with fully relativistic pseudopotentials.

  After converging the Kohn-Sham electronic structure, the
  valence and low-lying conduction Bloch bands were
  converted into projected Wannier functions using the {\tt Wannier90}
  code package.\cite{wannier90} For MoS$_2$ the projected Wannier
  orbitals were constructed using sulfur $p$ states and Mo $d$
  states. The sulfur $s$ states were included in the \textit{ab
    initio} calculations, but the low lying $s$-like Bloch bands were
  excluded from the wannierization. Unoccupied states were included by
  disentangling \cite{souza_disentangling} bands up to 3.0 eV above the
  conduction band minimum. Finally the Wannier functions were used to
  construct the Kohn-Sham Hamiltonian in a tight-binding basis
  $H_{ij}(\mathbf{R})$, where $i,j$ denotes orbital indices within the
  unit cell and $\mathbf{R}$ is a lattice vector. The set of lattice
  vectors included were defined by the Wigner-Seitz cell corresponding
  to the applied \textit{ab initio} $k$-point mesh. For example, in
  pristine MoS$_2$ with a $8\times8$ $k$-point mesh, we have 22
  orbitals (Mo $d$ and S $p$) and 64 lattice vectors (some of which
  are equivalent).

%\tnewpage
  \subsection{Tight-binding calculations}\label{tb}

%
%%  \ism{We need an opening sentence making a bridge between the {\it ab
%      initio} methods of the previous section and what is done
%    here. That's also why I expanded the title, otherwise a reader
%    might think that the paper contains a mix of {\it ab initio}
%    calculations on the one hand, and conventional TB calculations on
%    the other, with little relation between the two. In order to make contact between the two, we should say something like:\\
%    {\it Now that now that we have mapped the low-energy electronic
%      structure onto a Wannier basis, we can use it as an ``exact''
%      tight-binding basis to calculate efficiently and accurately
%      various quantities in $k$-space.} Then we should refer to
%    Ref.~\onlinecite{souza_disentangling} regarding the interpolation
%    of the energy bands, and to Ref.~\onlinecite{wang-prb06} for the
 %   Berry curvature, explaining which matrix elements are needed in
 %   each case.}
The majority of calculations in present work are tight-binding calculations
with parameters obtained from a Wannier representation of the Kohn-Sham Hamiltonian 
$H_{ij}(\mathbf{R})$. In a tight-binding framework, the calculation of band structures 
from $H_{ij}(\mathbf{R})$ is of course
equivalent to the standard Wannier interpolation.\cite{souza_disentangling} At the sampled set of 
$k$-points, $H_{ij}(\mathbf{R})$ will thus yield
the calculated Kohn-Sham eigenvalues and between the sampled points it smoothly interpolates.
Similarly, a rigorous Wannier interpolation scheme can be constructed for 
the AHC,\cite{wang_ahc} but this quantity cannot be calculated exactly in a 
bare tight-binding framework since the information contained in $H_{ij}(\mathbf{R})$ is not
enough to evaluate the Berry curvature. 

To see this explicitly we
will briefly state the relevant expressions below. The starting point is the Berry curvature
in its spectral representation where it can be written
\begin{align}\label{berry_tb}
 \Omega_{\alpha\beta}(\mathbf{k})&=\sum_{m,n}(f_{n\mathbf{k}}-f_{m\mathbf{k}})\\
 &\times\frac{\langle u_{m\mathbf{k}}|\nabla_\alpha H(\mathbf{k})|u_{n\mathbf{k}}\rangle\langle u_{n\mathbf{k}}|\nabla_\beta H(\mathbf{k})|u_{m\mathbf{k}}\rangle}{(\varepsilon_{n\mathbf{k}}-\varepsilon_{m\mathbf{k}})^2},\notag
\end{align}
with $\nabla_\alpha\equiv\partial/\partial k_\alpha$. We let
$\varphi_i$ denote at set of localized orbitals and expand the Bloch states
as
\begin{align}\label{eq:psi}
|\psi_{n\mathbf{k}}\rangle=\sum_i{C_{ni\mathbf{k}}}|\chi_{i\mathbf{k}}\rangle=\sum_{i\mathbf{R}}C_{ni\mathbf{k}}e^{i\mathbf{k}\cdot(\mathbf{R}+\mathbf{t}_i)}|\varphi_{i\mathbf{R}}\rangle
\end{align}
where
$\mathbf{t}_i=\langle\phi_{i\mathbf{0}}|\hat{\mathbf{r}}|\phi_{i\mathbf{0}}\rangle$. Note
that the inclusion of $\mathbf{t}_i$ is purely a matter of
convention. However, the present convention will prove highly
convenient below. Matrix elements of the Bloch Hamiltonian and their gradients can now be written as
\begin{align}
 &H_{ij\mathbf{k}}=\langle\chi_{i\mathbf{k}}|\hat H|\chi_{j\mathbf{k}}\rangle=\sum_\mathbf{R}e^{i\mathbf{k}\cdot(\mathbf{R}-\mathbf{t}_i+\mathbf{t}_j)}H_{ij}(\mathbf{R}),\\
 &\nabla_\mathbf{k}H_{ij\mathbf{k}}=i\sum_\mathbf{R}(\mathbf{R}-\mathbf{t}_i+\mathbf{t}_j)e^{i\mathbf{k}\cdot(\mathbf{R}-\mathbf{t}_i+\mathbf{t}_j)}H_{ij}(\mathbf{R})
\end{align}
%\tnewpage
%
%\ism{This sounds too negative, it's just the nature of the problem,
%  and again this is well-understood and thoroughly discussed in the
%  literature, namely Ref.~\onlinecite{wang-prb06}. For the diagonal
%  approximation for Berry-like quantities, we could again refer to the
%  above {\tt pythtb} notes.}
%
%The matrix elements appearing in Eq. \eqref{berry_tb}
%cannot be obtained from the alone. In particular we have
and in terms of these, the matrix elements appearing in Eq. \eqref{berry_tb} becomes
\begin{widetext}
\begin{align}
 J_{\alpha mn}(\mathbf{k})&\equiv\langle u_{m\mathbf{k}}|\nabla_\alpha H(\mathbf{k})|u_{n\mathbf{k}}\rangle\notag\\
 &=\sum_{ij\mathbf{R}}C^*_{mi\mathbf{k}}C_{nj\mathbf{k}}e^{i\mathbf{k}\cdot(\mathbf{R}-\mathbf{t}_i+\mathbf{t}_j)}\langle\varphi_{i\mathbf{0}}|e^{i\mathbf{k}\cdot\hat{\mathbf{r}}}\nabla_\alpha H(\mathbf{k})e^{-i\mathbf{k}\cdot\hat{\mathbf{r}}}|\varphi_{j\mathbf{R}}\rangle\notag\\
 &=-i\sum_{ij\mathbf{R}}C^*_{mi\mathbf{k}}C_{nj\mathbf{k}}e^{i\mathbf{k}\cdot(\mathbf{R}-\mathbf{t}_i+\mathbf{t}_j)}\langle\varphi_{i\mathbf{0}}|[\hat{\mathbf{r}},\hat H]|\varphi_{j\mathbf{R}}\rangle\notag\\
 &=-i\sum_{ijl\mathbf{R}\mathbf{R}'}C^*_{mi\mathbf{k}}C_{nj\mathbf{k}}e^{i\mathbf{k}\cdot(\mathbf{R}-\mathbf{t}_i+\mathbf{t}_j)}
 [\langle\varphi_{i\mathbf{0}}|\hat{\mathbf{r}}|\varphi_{l\mathbf{R}'}\rangle\langle\varphi_{l\mathbf{R}'}|\hat H|\varphi_{j\mathbf{R}}\rangle-\langle\varphi_{i\mathbf{0}}|\hat H|\varphi_{l\mathbf{R}'}\rangle\langle\varphi_{l\mathbf{R}'}|\hat{\mathbf{r}}|\varphi_{j\mathbf{R}}\rangle]\notag\\
 &=-i\sum_{ijl\mathbf{R}\mathbf{R}'}C^*_{mi\mathbf{k}}C_{nj\mathbf{k}}e^{i\mathbf{k}\cdot(\mathbf{R}+\mathbf{R}'-\mathbf{t}_i+\mathbf{t}_j)}
 [\mathbf{r}_{il\mathbf{R}'}H_{lj\mathbf{R}}-H_{il\mathbf{R}'}(\mathbf{r}_{lj\mathbf{R}}+\mathbf{R}'\delta_{\mathbf{R}\mathbf{0}}\delta_{lj})]\notag\\
 &=-i\sum_{ijl}C^*_{mi\mathbf{k}}C_{nj\mathbf{k}}
 [\mathbf{r}_{il\mathbf{k}}H_{lj\mathbf{k}}-H_{il\mathbf{k}}\mathbf{r}_{lj\mathbf{k}}]+i\sum_{ij\mathbf{R}}C^*_{mi\mathbf{k}}C_{nj\mathbf{k}}\mathbf{R}e^{i\mathbf{k}\cdot(\mathbf{R}-\mathbf{t}_i+\mathbf{t}_j)}H_{ij\mathbf{R}}\notag\\
 &=\sum_{ij}C^*_{mi\mathbf{k}}C_{nj\mathbf{k}}\nabla_\alpha H_{ij\mathbf{k}}+i\sum_{ijl}C^*_{mi\mathbf{k}}C_{nj\mathbf{k}}
[(\mathbf{t}_i\delta_{il}-\mathbf{r}_{il\mathbf{k}})H_{lj\mathbf{k}}-H_{il\mathbf{k}}(\mathbf{t}_l\delta_{lj}-\mathbf{r}_{lj\mathbf{k}})].
\end{align}
\end{widetext}
In the present work we have made the diagonal approximation where
$\mathbf{r}_{ij\mathbf{R}}=\delta_\mathbf{R0}\delta_{ij}\mathbf{t}_i$
and we thus use $J_{\alpha
  mn}(\mathbf{k})=\sum_{ij}C^*_{mi\mathbf{k}}C_{nj\mathbf{k}}\nabla_\alpha
H_{ij\mathbf{k}}$. With this approximation, the problem can be mapped exactly
to a tight-binding calculation with parameters obtained from the Kohn-Sham Hamiltonian
in a basis of Wannier functions. This turns out to be an excellent
approximation. The Berry curvature calculated with this method
cannot be distinguished with the naked eye from \textit{ab initio} calculations
(obtained by Wannier interpolation) and anomalous Hall conductivities calculated within the diagonal
approximation differ from \textit{ab initio} results 
by less than 2\%.

\subsection{Mapping the supercell onto the normal cell}
Unfolding band structures and curvatures involve the calculation of
matrix elements between pristine and supercell systems. A localized 
basis set allow us to perform the unfolding without direct reference to 
the pristine system.\cite{ku_unfolding}
\begin{figure*}[t]
    \includegraphics[width=8.0 cm]{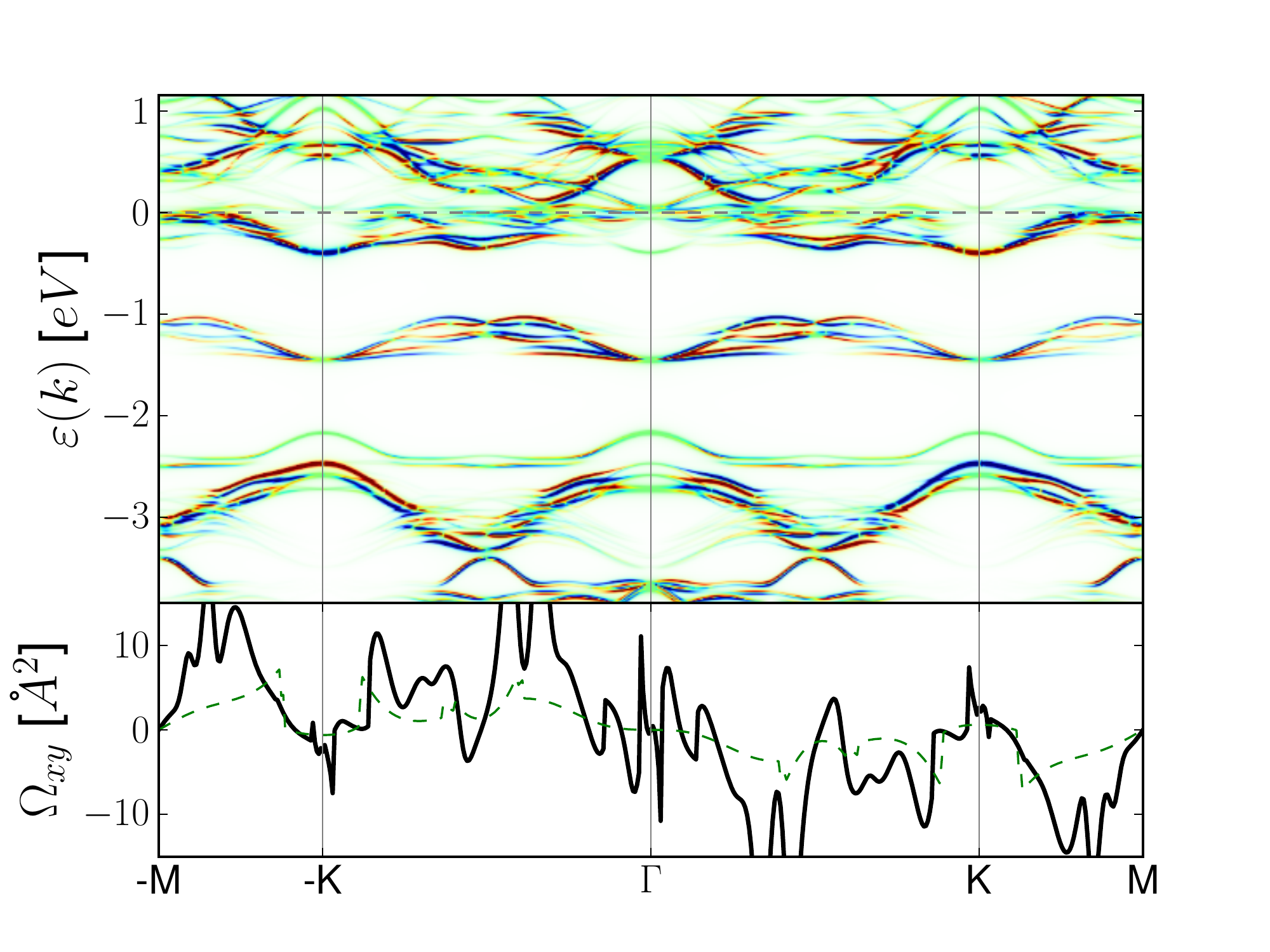}
    \includegraphics[width=8.0 cm]{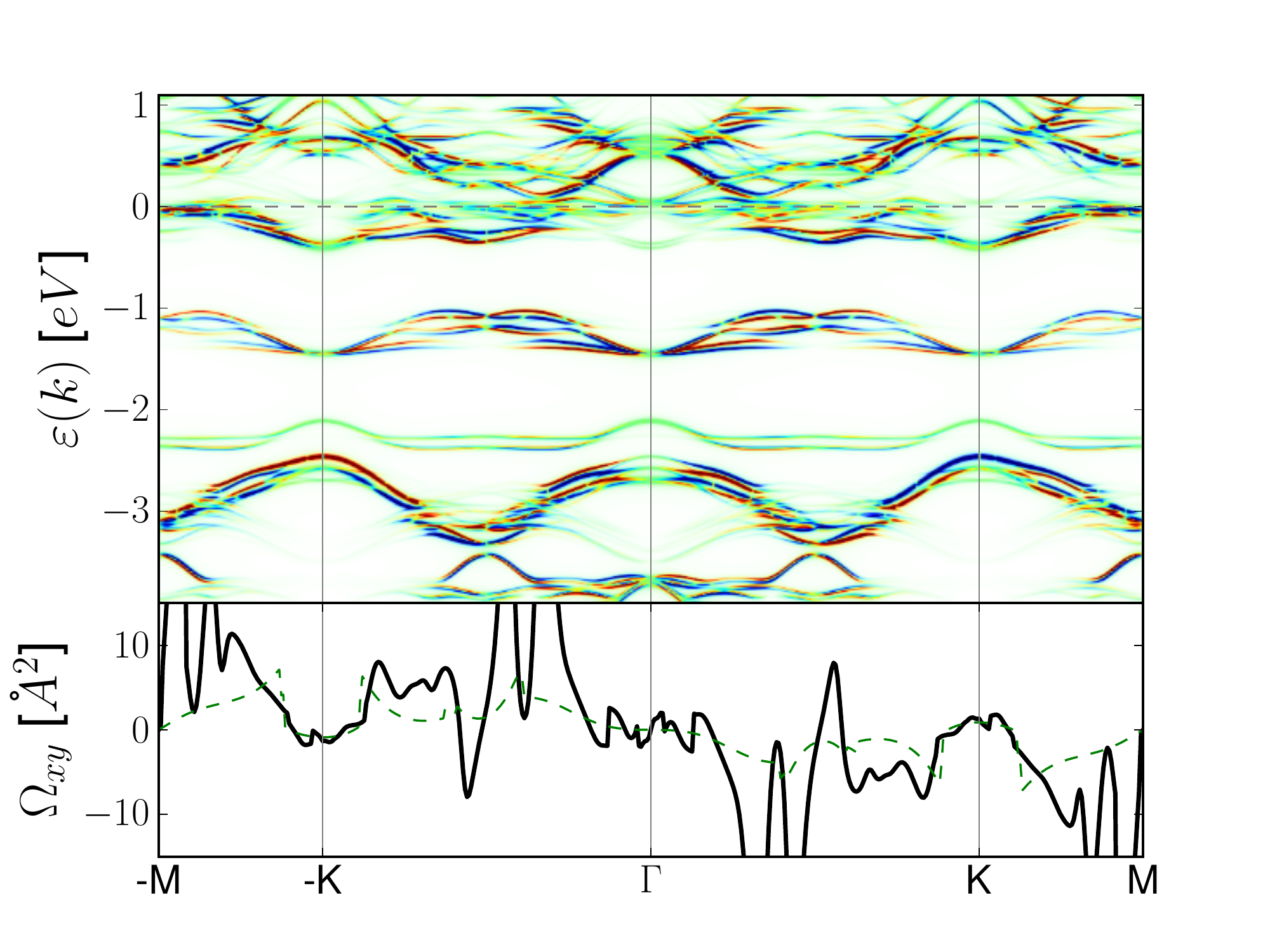}
    \caption{(Color online). Unfolded band structure and Berry
      curvature of MoS$_2$ in a $3\times3$ supercell with two sulfur
      vacancies at next nearest neighbor sites. Left: \textit{ab
        initio} calculation. Right: calculation from effective
      Hamiltonian constructed from a pristine $3\times3$ supercell
      with impurities added in the form of an influence Hamiltonian
      obtained from a single impurity calculation. The dashed lines show the
      Berry curvature of pristine MoS$_2$ with the Fermi level at the indicated
      position.}
\label{fig:effective_test}
\end{figure*}

We denote
pristine band, orbital indices, and crystal momentum by $n, i, \mathbf{k}$
respectively and supercell band, orbital, and crystal momentum
indices by $N,I,\mathbf{K}$ respectively. For the present purpose we will use
$\mathbf{r}$ for pristine lattice vectors and $\mathbf{R}$ for supercell
lattice vectors. We can thus consider the matrix element
\begin{align}
\langle\psi_{n\mathbf{k}}|\psi_{N\mathbf{K}}\rangle=\sum_{iI\mathbf{r}\mathbf{R}}C^*_{n	i\mathbf{k}}C_{NI\mathbf{K}}e^{-i\mathbf{k}\cdot(\mathbf{r}+\mathbf{t}_i)+i\mathbf{K}\cdot(\mathbf{R}+\mathbf{t}_I)}\langle\varphi_{i\mathbf{r}}|\varphi_{I\mathbf{R}}\rangle
\end{align}
Following Ku {\it et al.}\cite{ku_unfolding} we introduce a map that
uniquely identifies orbitals in the supercell with corresponding orbitals
in the pristine system. The map thus takes $I\rightarrow
\mathbf{r}'(I), i'(I)$ and we have
\begin{align}
\langle\varphi_{i\mathbf{k}}|\varphi_{I\mathbf{R}}\rangle=\langle\varphi_{i\mathbf{r}}|\varphi_{i'(I)\mathbf{R}+\mathbf{r}'(I)}\rangle=\delta_{ii'(I)}\delta_{\mathbf{r}\mathbf{R}+\mathbf{r'}(I)}.
\end{align}
The matrix element can then be written as
\begin{align}
\langle\psi_{n\mathbf{k}}|\psi_{N\mathbf{K}}\rangle&=\sum_{I\mathbf{R}}C^*_{ni'(I)\mathbf{k}}C_{NI\mathbf{K}}e^{-i\mathbf{k}\cdot[\mathbf{r}'(I)+\mathbf{R}+\mathbf{t}_{'(I)}]+i\mathbf{K}\cdot[\mathbf{R}+\mathbf{t}_I]}\notag\\
&=\sum_{I}C^*_{ni'(I)\mathbf{k}}C_{NI\mathbf{K}}e^{-\mathbf{k}\cdot[\mathbf{r}'(I)+\mathbf{t}_{i'(I)}]+i\mathbf{K}\cdot\mathbf{t}_I}
\delta_{\mathbf{K}[\mathbf{k}]},
\end{align}
where $[\mathbf{k}]$ is the set of crystal momenta that downfolds to
$\mathbf{K}$.

The idea of a map allows one to only work with the supercell 
system and avoid the explicit calculation of overlap
matrices. However, the procedure does require that the supercell
system considered is somewhat similar to the pristine reference system
and becomes ill-defined if there is not a unique way of relating
orbitals in the two systems. Furthermore, it is important to construct
the tight-binding Hamiltonian from projected Wannier functions as
opposed to maximally localized Wannier functions, since the latter may
differ significantly in otherwise similar systems.

In Figs. \ref{fig:MoS2_33} and \ref{fig:MoS2_33_shifted} we showed
examples of the band structures and Berry curvature of MoS$_2$ in a
$3\times3$ unit cell with a single sulfur vacancy, unfolded
onto the normal BZ.

\section{Effective Hamiltonians}\label{effective}
Here we briefly summarize the construction of effective Hamiltonians
as proposed in Ref.~\onlinecite{berlijn_effective}. In a tight binding framework, 
the effective Hamiltonian with $N$ impurities is constructed as
\begin{align}
 H^{Eff}_{IJ}&(\mathbf{R})=H^{SC}_{IJ}(\mathbf{R})\\
 +&\sum_{s=1}^NP(I,J,\mathbf{r}_s,\mathbf{R})H'_{i'(I)i'(J)}(\mathbf{r}'(I)-\mathbf{r}_s,\mathbf{r}'(J)+\mathbf{R}-\mathbf{r}_s)\notag,
\end{align}
where $I,J,\mathbf{R}$ denotes a supercell orbitals and lattice
vectors respectively. $\mathbf{r}$ is a normal cell lattice vector and
$i', \mathbf{r}'$ are maps from the supercell orbitals and lattice to
the normal cell. $\mathbf{r}_s$ denotes the position of impurity
$s$. The influence Hamiltonian is given by
\begin{align}\label{influence}
 H'_{ij}(\mathbf{r}_1,\mathbf{r}_2)=H^{Imp}_{ij}(\mathbf{r}_1,\mathbf{r}_2) - H^0_{ij}(\mathbf{r}_2-\mathbf{r}_1),
\end{align}
where $H^{Imp}_{i, j}(\mathbf{r}_1,\mathbf{r}_2)$ is constructed with
a map from the impurity to the normal cell. Since this is expressed in
a basis of normal cell lattice vectors $H^{Imp}$ and therefore $H'$ is
not periodic in simultaneous translations of $\mathbf{r}_1$ and
$\mathbf{r}_2$. We have used the partition function of Liu and
Vanderbilt\cite{liu_effective}
\begin{align}
P(I,J,\mathbf{r}_s,\mathbf{R})&=e^{-(d/d_c)^8}\\
d&=(|\mathbf{r}'(I)-\mathbf{r}_s| + |\mathbf{r}'(J)+\mathbf{R}-\mathbf{r}_s|)/2.\notag
\end{align}
with $d_c=9.0$ \AA.

For applications of the method we have performed \textit{ab initio}
calculations of the pristine and impurity systems and constructed
$H^{imp}$ and $H^0$ using projected Wannier functions. $H^{SC}$ is
then constructed as a straightforward repetition of $H^0$.

As a non-trivial test of the method we have performed an \textit{ab
  initio} calculation of a $3\times3$ unit cell of MoS$_2$ and with
sulfur vacancies at two next-nearest neighbor sites. We have then
constructed the same system with from the effective Hamiltonian: first we
construct obtain the tight binding model of a $3\times3$ unit cell of
MoS$_2$ by repeating the tight-binding Hamiltonian obtained from a
calculations of pristine MoS$_2$ (normal unit cell). We have then
constructed the influence Hamiltonian \eqref{influence} from the
pristine calculation and a calculation of MoS$_2$ in a $3\times3$ unit
cell with a \textit{single} sulfur vacancy. The influence Hamiltonian
is then added to the $3\times3$ tight-binding Hamiltonian at the two
nearest neighbor sites to obtain the effective Hamiltonian of a system
with two impurities. The unfolded bands and curvature of \textit{ab
  initio} and effective Hamiltonian calculations are shown in
Fig. \ref{fig:effective_test}. The unfolded spectral function is
nearly indistinguishable in the two case. However, the unfolded Berry
curvature is highly sensitive to the exact positions of bands near
avoided crossings and therefore exhibits larger deviations in the two
methods. Nevertheless, the Berry curvature obtained from the effective
Hamiltonian reproduces the main qualitative features (for example a
vanishing contribution in the vicinity of $K$ and $-K$) and we believe
that the configurational averaged Berry curvature obtained from the
effective Hamiltonian methods provides a reliable measure of the
effects of disorder on the curvature and VHC.

%\bibliography{bibfile}{}

%

\end{document}